\documentclass[final,5p,times,twocolumn]{elsarticle}
\usepackage{amssymb,amstext}
\usepackage[active]{srcltx}
\usepackage [autostyle, english = american]{csquotes}
\MakeOuterQuote{"}
\usepackage{graphicx}
\usepackage{dcolumn}
\usepackage{xfrac}
\usepackage{subfigure}
\usepackage{siunitx}
\usepackage[fleqn]{amsmath}
\usepackage{cases}
\newcommand\al{\textit{et al.}}

\pdfoutput=1

\begin{document}
\title{Complete characterization of sink-strengths for 1D to 3D mobilities of defect clusters: Bridging between limiting cases with effective sink-strengths calculations.}
\author[EDF]{Gilles Adjanor}
\ead{gilles.adjanor@edf.fr}
\address[EDF]{EDF Lab Les Renardi\`eres, Materials and Mechanics of Components Department, Moret-sur-Loing France}

\date{\today}

\begin{abstract}
In a companion paper, new analytical expressions of the cluster sink-strength (CSS) for two one-dimensional diffusers were proposed. These expressions are indispensable to any complete parameterisation of rate-equations cluster dynamics intending to model mobile interstitial clusters such as dislocation loops. Indeed, simulating the long-term microstructural evolution in systems with fast diffusing species such as self-interstitial atom (SIA) clusters relies on establishing the CSS for rate-equation cluster dynamics which critically depends on the complex mobilities of clusters.
In this second paper, we will estimate effective CSSs by object kinetic Monte-Carlo (OKMC) on wide ranges of radii, rotation energies, diffusion coefficients, and concentrations of both reaction partners. The symmetric roles of some parameters are used to infer a generic form of a semi-analytical expression of the CSS depending on all these interaction parameters. It is composed of, on one hand, the various analytical expressions established as limiting cases and, on the other hand, of fitted transition functions that allow a gradual switching between them. The analysis of the residuals shows that the overall agreement is reasonably good: it is only in the transition zones that some significant discrepancies are located. 
This semi-analytical expression answers our practical need for CSSs evaluation in our typical range of conditions, but further extending this range to much smaller diffusion coefficients ratios, it is quite striking to see that the domain for 1D-1D mobility is very extended. It is only when approaching a ratio of $10^{-6}$ that the 1D-0 CSSs start becoming more relevant than the adapted 1D-1D CSSs which have completely different orders of magnitude and kinetic orders. Finally, the semi-analytical CSS expression is validated by comparing its implementation in rate-equation cluster dynamics to OKMC simulations of interstitial clusters aggregation.
Companion paper: \cite{Adjanor1}
\end{abstract}

\begin{graphicalabstract}
\includegraphics[width=1.0\textwidth]{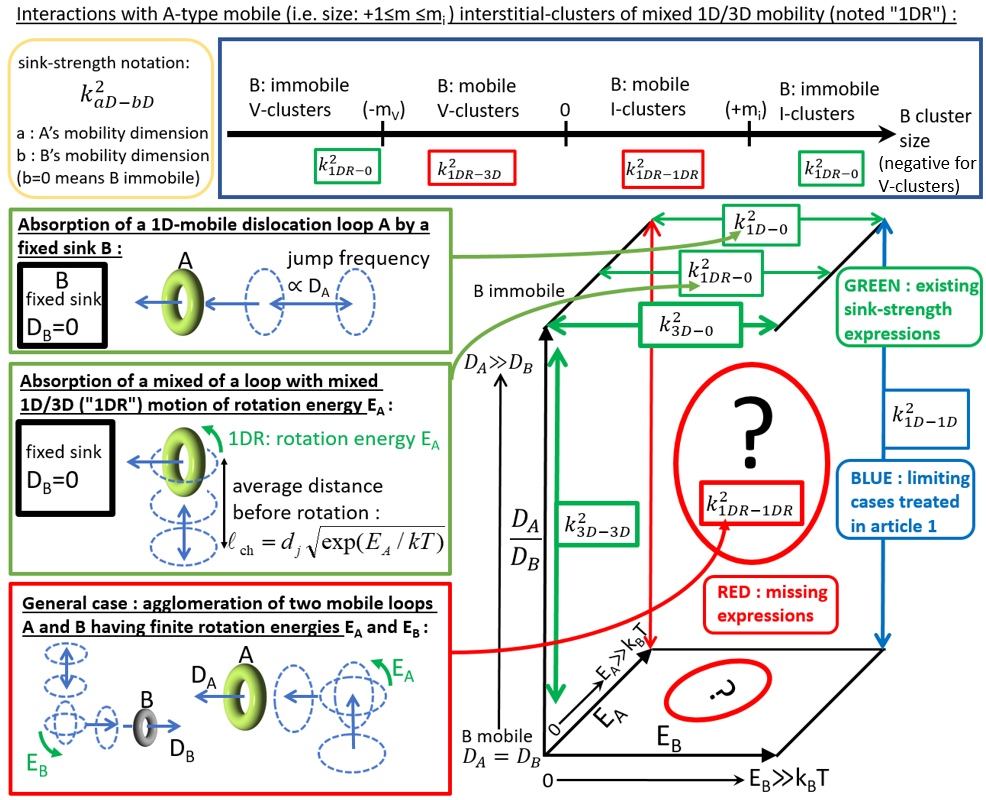}
\end{graphicalabstract}

\begin{highlights}
\item For interactions involving 1D-mobile clusters, most sink-strength expressions are missing which prevents the prediction of dislocation loops growth with rate-equations
\item When both interacting species have a mixed 1D/3D mobility (noted here "1DR"), the effective sink-strength can still be calculated with kinetic Monte-Carlo simulations
\item A general formula for these sink-strengths is proposed by bridging between the limiting cases established in the companion paper with a fit of the Monte-Carlo estimates
\item These general expressions are also well-validated once implemented in rate-equation cluster dynamics and then compared to the time consuming kinetic Monte-Carlo simulations
\end{highlights}

\maketitle

\section{Introduction}
Upon irradiation, the physical properties of fission and fusion reactors' materials evolve. This is primarily due to point defects and their clusters which are produced in displacement cascades, right after the initial neutron-atom high energy collision. In displacement cascades, the small fraction of point defects agglomerating into clusters may have a huge impact on the overall microstructural evolution: analytical models based on the "production-bias" \cite{WooPBM,SinghPBM} have shown the importance of the size distribution of the produced clusters rather than only considering the total number of monomers generated. This influence is even greater when considering the substantial mobility of defect clusters: basing on density functional theory or molecular dynamics simulations (MD), many studies \cite{Osetsky,Terentyev} showed that even large clusters of self-interstitial atom (SIA) are expected to be very mobile, possibly on equal footing with single-SIA's mobility. Depending on the considered system, the single-SIA should be either a dumbbell or a crowdion, while its clusters should be bundles of the latter, which are better described as dislocation loops are large sizes. At intermediate sizes, symmetries allow for a variety of low-energy cluster arrangements \cite{Marinica,Swinburne} but the most mobile ones are still describable as simple small loops \cite{Osetsky,Terentyev}. While large loops were confirmed experimentally \cite{Arakawa} to undergo a 1D random walk along their associated glide cylinder, small loops are rather expected to have a "mixed 1D to 3D" mobility \cite{Heinisch,Barashev,Trinkaus2002}. This intermediate case is physically understood as the thermally activated rotation of the loops' Burgers vector, so their actual trajectory consists of segments of 1D trajectories spanning the 3D space according to the crystallographic variants of the Burgers vector. This was observed in MD \cite{Soneda}, and Heinisch, Barashev, Trinkaus and co-workers \cite{Heinisch,Barashev,Trinkaus2002} proposed cluster sink-strength expressions necessary to account for the kinetics of such mixed mobility clusters being absorbed by a fixed spherical cluster. The related absorption rate is noted hereafter $"1DR-0"$, standing for "1D random walk with random rotations of the glide direction (1DR) with respect to (-) a fixed sink (0)". With this formalism, the rotation energy allows to describe all mixed mobilities between pure 3D-mobility (zero rotation energy) to pure 1D-mobility (very large rotation energy compared to $k_{B}T$, $k_{B}$ being Boltzmann's constant, and $T$ the temperature). In practice for systems like BCC-iron, the rotation energies range from zero to more than $2 \ \si{eV}$ \cite{chiapetto2015nanostructure} when assigning rotation energies from the single $\langle111\rangle$ dumbbell to large clusters of several tens of monomers.
The $1DR-0$ absorption rates  from the "master-curve" approach proposed by Heinisch \al (or the related cluster sink-strength, abbreviated CSS) are necessary to any mean-field modeling of long-term evolution such as rate-equation cluster dynamics (RECD) but, since they only account for interactions with a fixed-sink, they are not sufficient. A complete parameterisation of cluster dynamics should also account for absorptions between mobile clusters, starting from the interactions of a cluster class with itself (class "self-interactions"). In the companion paper \cite{Adjanor1}, analytical expressions for the absorption rates of two pure 1D-diffusers of types $A$ and $B$ (noted $1D_{A}-1D_{B}$) were proposed, including the diffusion "anisotropy" analog case (diffusion coefficient ratio $D_{A}/D_{B}$ being different than one). 

Here, we will rely on CSS estimates from object kinetic Monte-Carlo (OKMC) for the most general case of $1DR_{A}-1DR_{B}$ CSSs with arbitrary rotation energies.
Estimates from simulations are required because no general analytical formula is at hand. This is done for a very large set of conditions in terms of D-ratio (${\mathcal D}=D_{B}/D_{A}$) and of rotation energies $(E_{A}$, $E_{B})$ couples. The evolution of the CSS with these three parameters is then rationalised by proposing a semi-analytical formula, matching the analytical formulas for the limiting cases of $3D_{A}-3D_{B}$, $1D_{A}-3D_{B}$ and $1D_{A}-1D_{B}$ CSSs and fitting the transition between them thanks to a combination of sigmoid-type functions. The fit shows a good ability to reproduce the effective CSSs over the many orders of magnitude over which they evolve. Initially fitted on one couple of concentrations $(C_{A}, C_{B})$, this semi-analytical formula also reproduces effective CSSs with similar levels of accuracy for several others couples of concentrations, thus assessing for its broad validity. 
The interpretation of the CSS evolution heavily relies on the D-ratio exponents: as highlighted in limiting cases when the D-ratio varies, the CSS mostly varies like $(D_{A}/D_{B})^\delta$, the exponent being characteristic of the dimensions of both mobilities. A map of effective $\delta$ values helps the interpretations and the precision of the semi-analytical formula heavily relies on its main trends. On this map, domains of different mobility dimensionality appear, and their characteristic exponents are connected with the limiting cases for $1D_{A}-1D_{B}$ and $1D_{A}-3D_{B}$ anisotropic analog of the CSS ($(D_{A} \ne D_{B}$). But in the limited range of D-ratios investigated for a high precision fitting, these domains aren't closed: even for quite small D-ratios, the fixed sink limit is not reached. At section \ref{Closure}, we will investigate the closure of these domains, i.e. the convergence of effective CSSs towards the analytical CSS expression with respect to a fixed sink when the D-ratio gets lower (i.e $\Delta=\log_{10}(D_A/D_B)$ becomes higher). As we shall see, it is only when reaching much smaller D-ratios than in the initial set that the slowest species can be considered as immobile regarding CSS expressions. As discussed at section \ref{Discussion} and validated at section \ref{sectionCD} by comparing RECD to extensive OKMC simulations of complete microstructural evolutions, this implies a quite broad relevance of the established general CSS expression, even when the effective mobilities of species are considerably lowered after trapping by impurities or elastic fields. A more general consequence of this, the prevalence of coagulation (termed here as growth by mutual mobility) over Ostwald ripening is also discussed in the context of loops growth kinetics in typical irradiation conditions.

\section{OKMC methods and reference parameterisation}  \label{sectionMethodAndResultsResults}

\subsection{Reference OKMC parameterisation} \label{OKMCparam}
According to transmission electron microscopy (TEM) observations \cite{Arakawa,Hamaoka2010,Hamaoka2013}, once they are detrapped, even large (here "large" should be understood as visible in conventional TEM) dislocation loops can have large mobilities.
Some state-of-the-art object kinetic Monte-Carlo (OKMC) parameterisations like those of Chiapetto, Malerba \textit{et al.} \cite{chiapetto2015nanostructure,Note1} were aimed at reproducing these experimental observations: the jump frequency of a loop "object" is essentially a decreasing function of the loop's size \cite{Dudarev} and is given by the product of the exponential of migration energy (which is comparable to the Peierls barrier as the mobility of large loops can be seen as the glide of a circular dislocation) and an attempt frequency. 
It is important to note here that, owing to their enhancement by trapping impurities considerations, the effective migration energies adopted here are fortunately well above $k_{B} T$ (which is about $0.05 \si{eV}$ at $T=573\ K$) at the temperatures of typical interest.
At variance, if one would consider migration energies for crowdion bundles in perfectly pure iron devoid of any impurity then, according to Osetsky \al \cite{Osetsky2000} or Terentyev \al \cite{Terentyev} among other groups, they would be below $k_{B} T$ at $T=573\ K$. This would be problematic since KMC relies on the usual "separation of time scales" assumption \cite{Kramers,Chandrasekhar,Woo2017} and the related Markov-chain of rare-events formalism. Having systematic correlations between consecutive jump directions due to slow thermalisation (as Terentyev \al and others clearly observed in MD \cite{Terentyev}) makes the application of the usual master-equation formalism underlying the KMC algorithm doubtful. So, one should be careful not to apply directly the results of this paper to these situations which are, anyway, beyond the scope of conventional KMC and RECD methods.

Given this jump model of table \ref{tableMalerba}, the evolution of SIA clusters' diffusion coefficients from the monomer to nanometric loops will span over several orders of magnitude. Even if small, mobilities of large loops can be crucial to reproduce loop coarsening kinetics which can be driven either by Ostwald type ripening \cite{Moll} or by coagulation of mutually mobile species (also termed as "coagulation" in the literature \cite{Ratke}). To model these phenomena with a RECD model, the basic ingredients are the absorption rates of all combinations of reacting species whose D-ratios span over a large range. Values of attempt frequencies $\nu$, migration energies $E^m$ and rotation energies $E$ for different interstitial cluster sizes (number of SIA monomers, $n$) from one typical parameterisation of Malerba \al are given in table \ref{tableMalerba}. This gives us a hint of the typical orders of magnitudes and variations amplitudes of these physical parameters for BCC iron.

\begin{table}
\resizebox{0.4\textwidth}{!}{
\begin{tabular}{|c|c|c|c|c|c|c|}
\hline
$n$          & $R$         &$\nu$           &  $E^m$           &$D$                   & $E$       & $E$            \\
$ $          &$(\si{nm})$       & $(\si{s^{-1}})$     &  $(\si{eV})$          &$(\si{cm^{2} s^{-1}})$     &  $(\si{eV})$   &  $k_{B} T,$        \\
$ $          &$ $       & $ $     &  $ $          &$ $     &  $ $   &  $ T\simeq573\ K$        \\

\hline
$1 $           & $0.516$   &$\num{8.07e13}$  &  $0.31$          &$\num{8.31e-3}$      &  $0  $    &  $0$         \\
$2 $           & $0.567$   &$\num{3.41e14}$  &  $0.42$          &$\num{3.51e-2}$      &  $0  $    &  $0$         \\
$3 $           & $0.606$   &$\num{1.17e13}$  &  $0.42$          &$\num{1.21e-3}$      &  $0.2$    &  $4$         \\
$4 $           & $0.639$   &$\num{1.19e13}$  &  $0.80$          &$\num{1.23e-3}$      &  $0.4$    &  $8$         \\
$5 $           & $0.668$   &$\num{1.56e12}$  &  $0.1$           &$\num{1.60e-4}$      &  $0.6$    &  $12$         \\
$6 $           & $0.682$   &$\num{1.71e12}$  &  $0.2$           &$\num{1.76e-4}$      &  $0.8$    &  $16$         \\
$7 $           & $0.705$   &$\num{1.71e12}$  &  $0.2$           &$\num{1.76e-4}$      &  $1.0$    &  $20$         \\
$8 $           & $0.725$   &$\num{1.53e12}$  &  $0.2$           &$\num{1.58e-4}$      &  $1.2$    &  $24$         \\
$9 $           & $0.745$   &$\num{1.39e12}$  &  $0.2$           &$\num{1.43e-4}$      &  $1.4$    &  $28$         \\
$10$           & $0.764$   &$\num{1.28e12}$  &  $0.2$           &$\num{1.32e-4}$      &  $1.6$    &  $32$         \\
$11$           & $0.781$   &$\num{1.19e12}$  &  $0.2$           &$\num{1.22e-4}$      &  $1.8$    &  $36$         \\
$12$           & $0.798$   &$\num{1.11e12}$  &  $0.2$           &$\num{1.14e-4}$      &  $2.0$    &  $40$         \\
$\vdots$       & $\vdots$  &$\vdots$         &  $0.2$           &$\vdots$             &  $2.0$    &  $40$         \\
$60$           & $1.28$    &$\num{2.45e11}$  &  $0.2$           &$\num{3.15e-5}$      &  $2.0$    &  $40$         \\

\hline                                                                     
\end{tabular}}                                                              
\caption{Mobility parameters for interstitial clusters of size $n$ (in monomer content) in one typical parameterisation proposed by Malerba \text{et al.} \cite{Note1,chiapetto2015nanostructure}: rotation energies $E$, attempt frequencies $\nu$ and migration energies $E^{m}$ related to the diffusion coefficient $D$ (note that the radii $R$ were given assuming clusters are spherical, but they are augmented by capture efficiencies and spontaneous recombination distances.)}\label{tableMalerba}  
\end{table}
It should be stressed out that to have a physically relevant parameterisation of the loops' mobility, it would be very important to model at the same time the trapping centers that significantly reduces the loop's effective diffusion coefficients, otherwise, all loops would probably annihilate at the surfaces and none would be visible. Typical traps are impurities or carbon-vacancy complexes in the referred parameterisation. Nevertheless, this feature of the original model is not accounted for in the present study as it focuses on cluster-cluster interactions and not traps. It wouldn't be a major difficulty to include them in a dedicated RECD parameterisation, once the CSSs are adequately parameterised for the dimensionality of both mobilities.


\subsection{General scheme for effective sink-strengths computations} \label{section_OKMC_description}
In this study, we will calculate large sets of effective CSSs with OKMC. Because the populations of clusters constantly fluctuate in time, this cannot be done by a direct OKMC simulation but rather with a specific procedure to operate and count individual pair reactions at constant concentrations $C_{A}$ and $C_{B}$. The OKMC parameterisation in table \ref{tableMalerba} for the time evolution of the clusters' distribution is merely a guideline for the relevant ranges of parameters to be explored.

In the case general case where both reaction partners $A$ and $B$ can be mobile, we expect non-trivial dependencies of absorption rates on the couple of concentrations $(C_{A}$, $C_{B})$ and substantial dependence on the diffusion coefficient ratio ${\mathcal D}=D_{B}/D_{A}$. The effective CSS calculations should then be done with large numbers of clusters in the simulation box to allow for more flexibility in the choice of concentrations and better physical accuracy of estimates.

Malerba \al \cite{Malerba} used the OKMC code LAKIMOCA \cite{DomainLakimoca} to compare effective CSSs to the $1DR-0$ analytical expressions from the so-called "master curve approach" \cite{Heinisch,Barashev,Trinkaus2002}. In the latter study, effective $1DR-0$ sink-strengths were estimated using a parameterisation for interstitial type objects performing 1D-jumps along one over the four variants of the glide directions of the $1/2\langle111\rangle$ family. Changes of the glide direction (rotations) were chosen among all other possible stochastic events proportionally to their respective probabilities, and so that the average distance covered between two rotations is $\ell_{ch}=d_j \sqrt{\exp(E/k_B T)}$ on average (where $d_j$ is the jump distance, $E$ is defined as the rotation energy). 

The effective CSS calculation then consists in placing one single sink (the $B$-type particule), then operating many OKMC jump events for the mobile cluster (the $A$-type particule), and gathering very large statistics to estimate the average number of jumps before their contact. Yet, for more general CSS estimations, such a procedure has some limitations: the concentration of both species are equal ($C_{A}=C_{B}=1/V$) so exclusion volume effects (see for instance Redner \al \cite{Redner,krapivsky2010kinetic}) between particles of the same type cannot be accounted for. On top of that, having a single sink in a periodic box, the resulting CSS corresponds to the case of a cubic mesh of sinks instead of a random distribution of them. Nevertheless, at least the first limitation, is not critical for the latter authors' purpose since there is no need to vary both concentrations when one of the reaction partners is immobile and the analytical form of the assessed absorption rate is known \textit{a priori}.

Oppositely, for the more general framework the present study without \textit{a priori} knowledge of the general CSS expressions, a new procedure for estimating effective CSSs must be established. The general scheme can be described as follows: 
\begin{enumerate}
\item One places $N_{A}=C_{A} V$ and $N_{B}=C_{B} V$ A and B species at random positions in the box of volume $V$, but away from reaction distances ($R=R_{A}+R_{B}$, $R_{AA}=2 R_{A}$, $R_{BB}=2 R_{B}$) of all other objects.
\item All clusters may jump sequentially according to the OKMC algorithm and to their mobility characteristics ($D_{A}$, $D_{B}$, $E_{A}$, $E_{B}$), until one object enters a reaction volume.
\item Once a heterotypic reaction (i.e. a $A-B$ reaction) occurs, the time elapsed since the previous reaction of this type is recorded. Then, one of the two species is moved to a random place of the box, away from all possible reactions' distances. This is necessary to keep the concentration of species constant while preventing overestimating absorption rates if the reacting defect pair would not be separated after the reaction time is recorded.
\item Once a homotypic reaction ($A-A$ or $B-B$ reactions) should occur, the associated time span is not recorded, and the reactions partners are randomly replaced away from any capture distance, as in the previous case. Without this precaution, the clusters capture volumes would overlap and the sink strength would be underestimated.
\item Periodically when, on average, each defect should have reacted a few times, all the clusters are randomly placed in the box again, thus allowing sampling of initial distributions of clusters whose effects can be especially important at low volume fractions of 1D-mobile species \cite{Redner}.
\end{enumerate}
This procedure shares some common points with that of Amino \al \cite{Amino} but it is meant to be more robust for the wide range of mobility parameters $S={(C_{A}, R_{A}, D_{A}, E_{A}, C_{B}, R_{B}, D_{B}, E_{B})}$ that we wish to explore in this paper.

When dealing with 1D mobilities, the convergence of OKMC's CSS estimations is known to be very slow  compared to CSS estimations for 3D diffusers (as already noted by Malerba \al \cite{Malerba} in the fixed sink case). This well-known fact \cite{Redner} stems from the tremendous inefficiency of the "recurrent" 1D diffusers to sample the 3D-space compared to the "transient" 3D diffusers. For the same reasons, it is shown in \ref{Search} that the conditions for a reliable convergence of object kinetic Monte-Carlo (OKMC) CSS estimates are extremely delicate and demanding. Nevertheless, thanks to validated simple models, the conditions for the estimates' convergence can be established using three criteria: 
\begin{enumerate}
\item the first criterion states that the number of the minority specie ($Min(N_A,N_B)$) should be greater than the smallest box dimension in unit cells (thus imposing very large boxes as $N \propto C L^3$), 
\item the second criterion states that the number of reactions to be performed should be at least equal to the total number of species $N_{A}+N_{B}$,
\item the last criterion is that a few tens of CSS estimates should be used to assess their standard deviations to mean ratio. 
\end{enumerate}
In the calculations presented in the rest of this article, the three preceding convergence criteria are met, so the computed CSSs are now referred to as "effective CSSs", $k^2_{\text{eff}}$, instead of estimates of the CSSs.

\subsection{The issue of establishing a general CSS expression for the $1DR-1DR$ case}

In a first step, the D-ratios used for CSS calculations will vary from ${\mathcal D}=D_{B}/D_{A}=10^{-3}$ to $1$. In table \ref{tableMalerba}, rotation energies values span from $0$ for an interstitial monomer which is here considered as fully 3D-mobile, to $2\ eV$ ($\simeq 40\ k_{B} T$ at $T=573\ {\si K}$) which, for not too low defect volume fractions and moderate temperatures, can be considered as sufficient to ensure a completely one-dimensional trajectory before absorption. This is also the case for temperature $T=573\ {\si K}$ considered in the present OKMC calculations. Problematically, we see that without \textit{a priori} assumptions on the dependence of CSSs on defect parameters for an arbitrary couple of values $(E_{A}, E_{B})$, the required number of calculations can be extremely large. Indeed, if we would mesh the parameter space with only six values for each of the eight dimensions $(C_{A}, R_{A}, D_{A}, E_{A}, C_{B}, R_{B}, D_{B}, E_{B})$, there would be millions of parameter combinations to consider. On top of that, for each of it, the CPU time required for convergence is very variable and its spans from minutes to weeks, even applying the above mentioned simulation setup rules for an optimal convergence. So simplifying assumptions are needed to limit the number of CSSs to estimate.

The first important assumption is that the radius dependencies can be cast into the single parameter $R=R_{A}+R_{B}$, as it appears in the previously established analytical CSS expressions from the companion paper. This approximation should be well verified when $R_{A} \simeq R_{B}$. If $R_{A} \ll R_{B}$ it can be questioned because, if homotypic reactions are prevented, then there will be a coexistence of zones with high concentrations of A-species with depleted ones while B-species will tend to be evenly spaced. As the simplifying assumption on the capture distances happens to be numerically well-validated for $3D-3D$, $1D-3D$ and $1D-1D$ CSSs (figures not shown for brevity) we may not question it much further in the general case.

\begin{figure}
\includegraphics[width=0.5\textwidth]{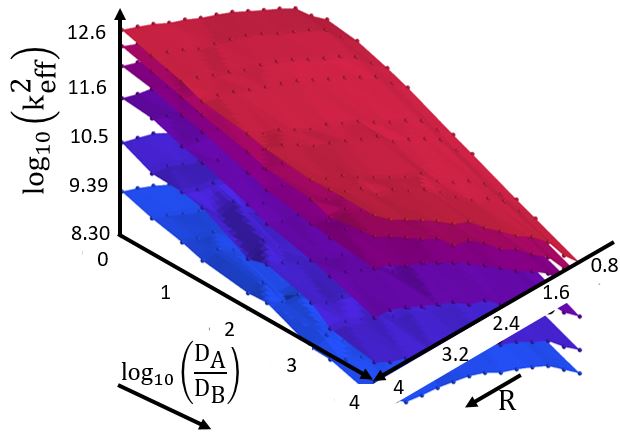}
\caption{The logarithm of the effective CSS, $k^2_{\text {eff}}$, in $\si{cm^{-2}}$ (estimated with the OKMC procedure \ref{section_OKMC_description} at $573 \ {\si K}$) as a function of the radius (in $\si{nm}$) and of $\Delta=\log_{10}(D_A/D_B)$ for the conditions $(E_A=40, E_B=40)\ (k_{B}T)$ and $C_A=C_B$ ranging from $\num{3.4e+16}$,$\num{6.8e+16}$, $  \num{1.7e+17}$, $\num{3.4e+17}$, $\num{5.1e+17}$, $\num{6.8e+17}$ $\si{cm^{-3}}$  (from blue to red surfaces, for the references to color in this figure, the reader is referred to the web version of this article). }
\label{k2_EA@2.0_EB@2.0}
\end{figure}

The figure \ref{k2_EA@2.0_EB@2.0} represents the effective CSS computed from the OKMC estimation procedure with $(E_{A}=40 k_{B}T,E_{B}=40 k_{B}T)$ at various concentrations. With both rotation energies as high as $40 k_{B}T$ at $573\ {\si K}$, the particles will have a very large diffusion length before rotation, so they can safely be treated as purely 1D-mobile and thus the comparison with the analytical $k^2_{1D-1D}$ is relevant. For each $R$ value, the linear relation between $\Delta=\log_{10}(D_{A}/D_{B})$ and $\log_{10}(k^2_{\text{eff}})$ in the figure indeed suggests a power-law dependency of the CSSs on the D-ratio. Detailed analysis indeed shows that it follows the expected characteristic exponent $-1/3$ expected from the diffusion anisotropy analogy highlighted in the companion paper. The evolution of CSS surfaces with $R$ are more complex, and changing the representation to $\log_{10}(R)$ would not reveal a simple and general scaling. Again, this is not surprising, since according to \cite{Adjanor1}:
\begin{equation}
k^2_{1D-1D} =  2\pi {\overline{R_{\text{eff}}}} \frac{4}{-\ln\left(\pi^{2}(C_{A}+C_{B})R^{3}/2\right)} C_{B}, \text{for\ } D_{A}=D_{B} \label{Eq_k2_1D-1D}
\end{equation}
the CSS should depend on $R$ through $R_{\text{eff}}=\frac{\pi}{4} R$ and the inverse logarithm of $R^3$.

From the preceding part, we can conclude that the proposed analytical expression matches well the computed effective CSS for the $1D-1D$ case. Other limiting cases like $1D-3D$ and $3D-3D$ (rotation energy couples set to $(E_{A}=0,E_{B}=40 k_{B}T)$ and $(E_{A}=0,E_{B}=0)$ respectively) were checked to follow their related limiting case CSS expressions, although, for the sake of conciseness, the corresponding graphs are not shown here. 

However, this approach reaches its limits when dealing with intermediate values of rotation energies $0<E<40\ k_{B}T$ that do not correspond to a well-defined limiting case: with a direct plot of a set of CSS surfaces depending on D-ratio and $R$, it can be difficult to determine which CSS analytical expression is the best match. Moreover, pursuing the approach of estimating CSSs for many different couples of concentrations, it may be impossible to rationalise the results: concentrations relevant to typical irradiation or nucleation conditions span over too many orders of magnitudes. We will thus adopt a different approach in the next section.

\section{Semi-analytical approach for general sink-strengths calculations} \label{semiAnalytical}
We will now focus on establishing a semi-analytical expression of CSSs for the general case of $(E_{A}, E_{B}) \in [0,\ 40]\times[0,\ 40]$ ($k_B T$) and reproducing the identified analytical expressions for limiting cases. Relying on the analytical expressions for the concentration dependencies will solve the CSS generalisation issue that was previously discussed. The functions allowing for the transition between the limiting cases will be fitted on simulation results which makes the approach not fully analytical but only semi-analytical. For this fitting procedure, a first set of $1728$ conditions has been simulated at $T=573 \ {\si K}$. It corresponds to the following parameters ranges: 
\begin{eqnarray} 
&&C_{A}=C_{B}=\num{1e17} \si{cm^{-3}}, \\ 
&&R_{A}=R_{B}=0.5\ \si{nm},\\ 
&&(E_{A}, E_{B}) \in \{0, 2, 4, 6, 8, 10, 12, 14, 16, 18, 20, 40\}^2 (k_{B}T),\\ 
&&D_{A}+D_{B}=\num{3.12e-5} \si{cm^{2} s^{-1}}, (D_{A} \ge D_{B})\\
&&\Delta=-\log_{10}({\mathcal D})=\log_{10}(D_{A}/D_{B}), \\
&&\Delta \in \left\{0, 0.27, 0.54, 0.81, 1.09, \right.\\
&&\left.1.36, 1.63, 1.9, 2.18, 2.45, 2.72, 3\right\}.
\end{eqnarray}
Note that these parameters have been chosen with some guidance from the parameterisation in table \ref{tableMalerba}, but that this choice does not impact the generality of the results to come, thanks to their semi-analytical character.
One advantage of having the condition $C_{A}=C_{B}$ fixed to a not too low density is that it eases the search for the optimal simulation parameters (such as internal variables for OKMC's "linked-cell" acceleration technique, box size, single run duration, etc.) valid for the whole set of conditions. Thus, we could obtain coefficients of variation ${\sigma(k^2_{\text {est}})}/{\overline{(k^2_{\text {est}})}}$ (where $\sigma$ is the standard deviation) of $1\%$ on average over all the $1728$ conditions and about $5\%$ in the worst cases. The average number of estimates for each condition (runs with different initial placements of clusters) is equal to ten.

\begin{figure}
\includegraphics[width=0.5\textwidth]{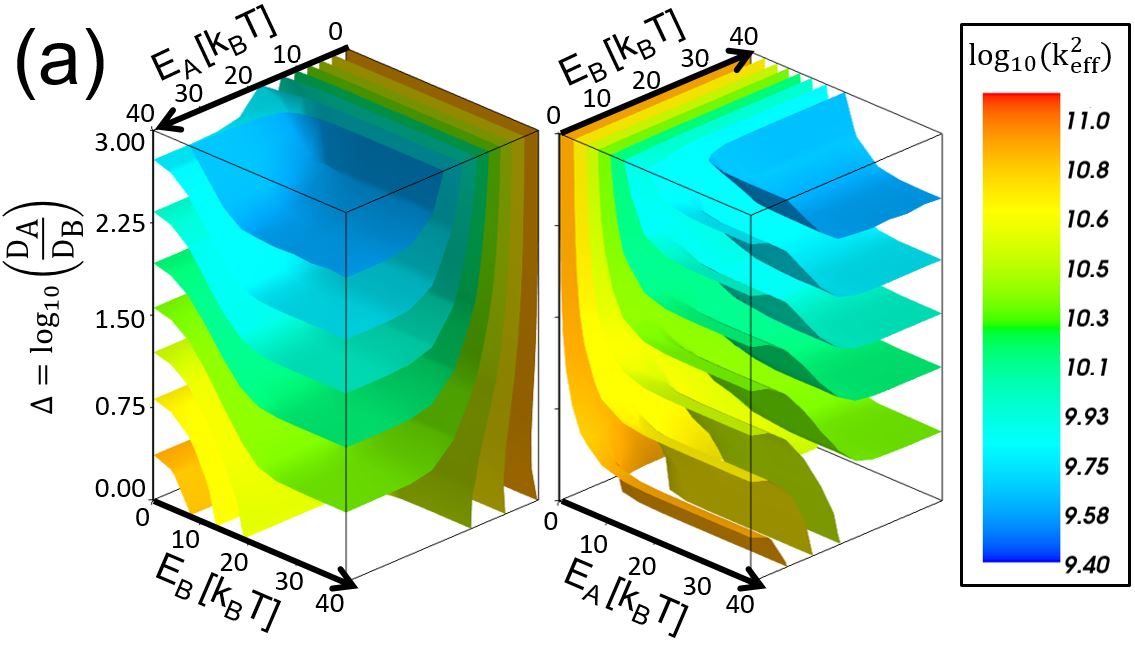}
\includegraphics[width=0.5\textwidth]{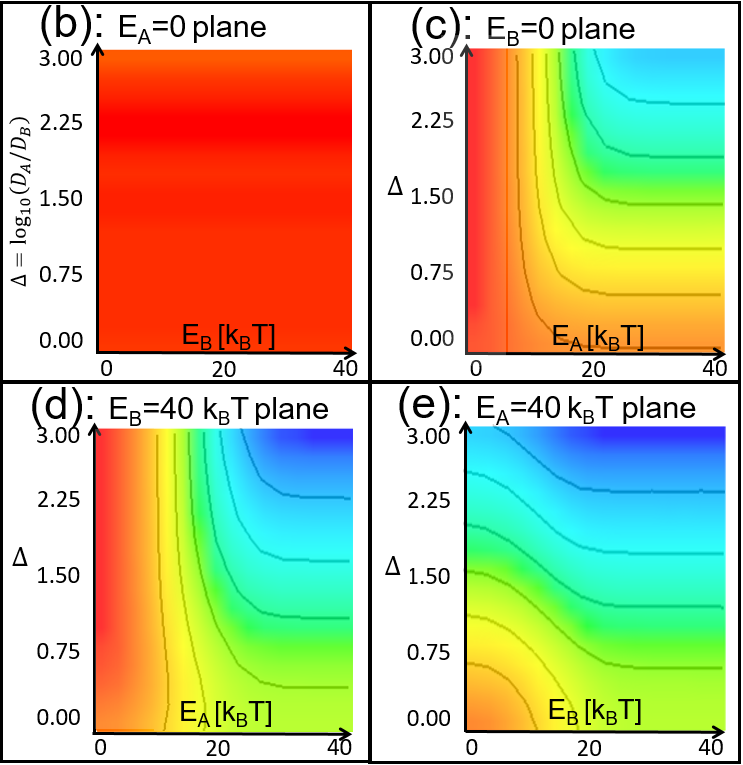}
\caption{(a): two rotated views of the same $\log_{10}(k^2_{\text{eff}})$ ($k^2_{\text{eff}}$ expressed in $\si{cm^{-2}}$) isosurfaces in the $(E_A, E_B, \Delta)$-space (represented using the Mayavi library \cite{Mayavi}). 
(b,c,d,e): plane projections of $\log_{10}(k^2_{\text{eff}})$ and isocontours (the grey curves in (c,d,e) panels) for guidelines.}
\label{cube1}
\end{figure}

The results are shown in the top panel (a) of Fig.~\ref{cube1} and the cross sections panels (b,c,d,e). They are displayed in the form of CSS-isosurfaces based on the $1728$ CSS estimates. The overall shape is quite complex, but we can nevertheless draw some trends. First, in Fig.~\ref{cube1}-(b) we see that for values of A species' rotation energy $E_{A}$ close to zero, the CSSs reach their highest values. This is also quite independent of both $E_{B}$ and the diffusion coefficients ratio: the corresponding isosurfaces are almost flat and parallel to the $E_{A}=0$ plane. This is because when $D_{A}$ is greater than $D_{B}$ and $A$ is pure 3D-diffuser, the type of mobility of $B$ has very little influence on absorption probability and the overwhelming efficiency of 3D-mobility will completely dominate the kinetics.

\begin{figure}
\includegraphics[width=0.5\textwidth]{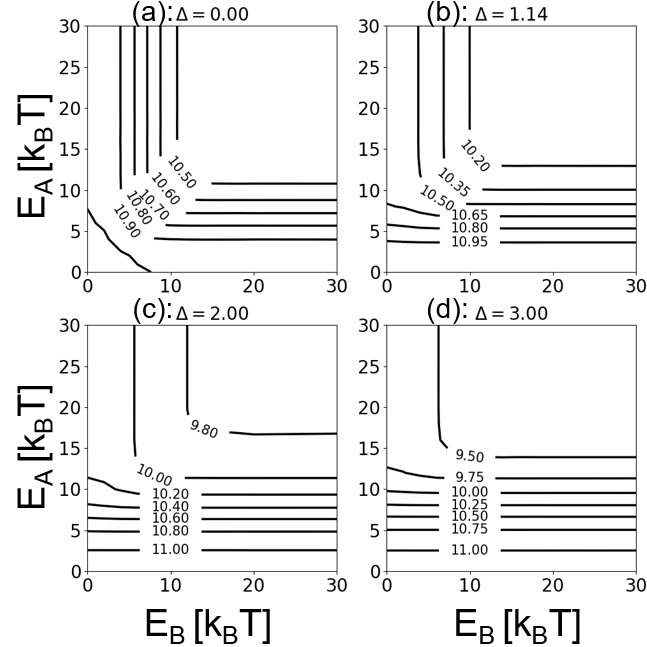}
\caption{Contours of the decimal logarithm of effective CSSs in $\si{cm^{-2}}$ in the $(E_A,E_B)$ planes: (a) for $D_A=D_B$ , and (b,c,d) for increasing diffusion coefficients ratios $\Delta$.}
\label{k2_DA-DA}
\end{figure}

To further interpret the variations of the CSS isosurface, we should first consider the horizontal cross section in Fig.~\ref{k2_DA-DA}-(a). This corresponds to $D_{B}$ is equal or very close to $D_{A}$ ($\Delta$ close to $0$). In this case, $E_{A}$ and $E_{B}$ should play symmetric roles for the CSS. This is indeed what is observed in Fig.~\ref{k2_DA-DA}-(a). The hyperbolic-like shapes of the contour lines can be simply interpreted by the fact that because $D_A=D_B$, for $E_A$ fixed, the CSS depends very weakly on $E_{B}$, so iso-CSS lines should be parallel to $E_{B}$ (and conversely for $E_B$ fixed). Also, we note that, globally, whatever the couple $(E_{A},E_{B})$, the CSS varies quite weakly in this "isotropic" diffusion analog case $D_A=D_B$ \cite{Adjanor1}. Nevertheless, it will be important to reproduce these hyperbolic-like shapes, because by extrusion and non-uniform shear along the $z$-axis (see the evolution of hyperboles in Fig.~\ref{k2_DA-DA}-(b,c,d)), they generate the complex isosurfaces of Fig.~\ref{cube1}-(a).

Indeed, another salient feature of Fig.~\ref{cube1}-(a,c,d,e) is that a series of isosurfaces parts are almost identical up to a constant translation along the $z$-axis. This is particularly striking for large values of $E_{A}$. In this semi-logarithmic representation, the constant spacing along $\Delta$ between isosurfaces is the signature of the power-law dependencies highlighted in the limiting cases of CSS expressions. This holds for specific parts of the $(E_A,E_B,\Delta)$-parameter space: the fact that isosurfaces reach plateau values, notably when both $E_{A}$ and $E_{B}$ are large, suggests a quite wide range of validity of limiting case analytical formulas. Between these plateau regions, there are very clear transition zones where isosurfaces have sigmoid shapes (most clearly seen in Fig.~\ref{cube1}-(e)) .

\begin{figure}
\includegraphics[width=0.5\textwidth]{cube-montage_NEW.jpg}
\includegraphics[width=0.5\textwidth]{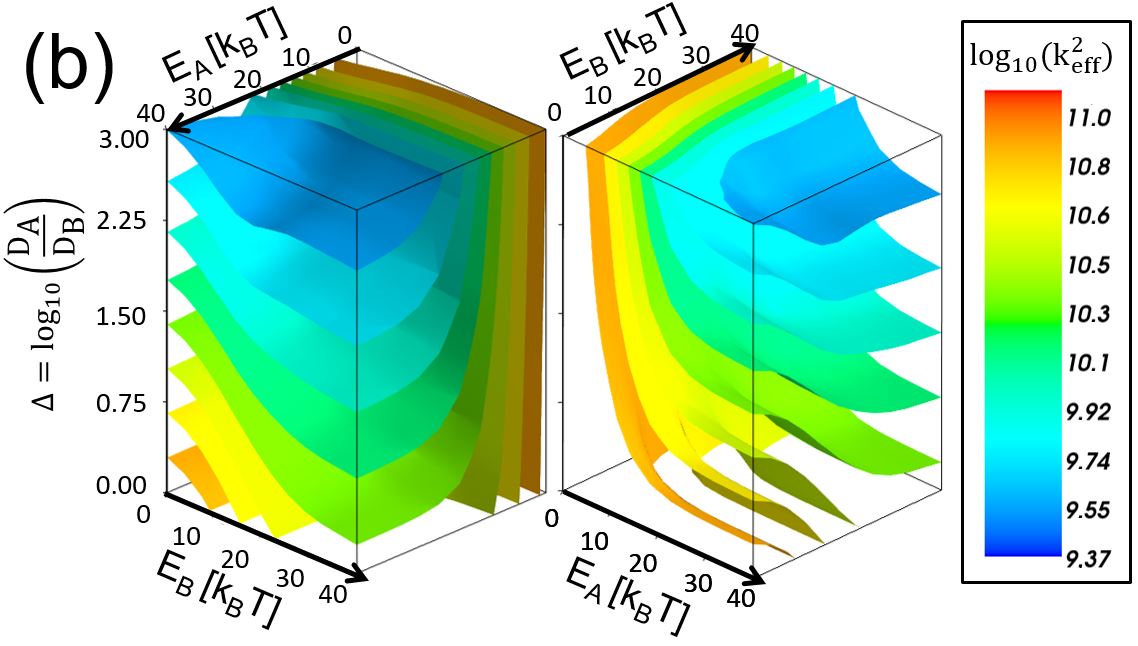}
\caption{(a) two rotated views of the same set of $1728$ CSS values estimated with OKMC at $T=573\ {\si K}$. Their logarithm $\log_{10}(k^2_{\text{eff}})$ (the CSS being in $\si{cm^{-2}}$) is represented as isosurfaces in the $(E_A, E_B, \Delta)$-space. (b) same views of isosurfaces obtained from the semi-analytical formula Eq. \ref{semiEmpirical} applied to the same set of conditions.}
\label{cube2}
\end{figure}

All these observations will now guide us to build a semi-analytical CSS expression for any couple of rotation energies, whose "transition" functions' coefficients will be adjusted on the present data set. Hereafter, we will use the well known "sigmoid" function (also known as the logistic function):
\begin{equation}
\sigma(\lambda, \varepsilon, x)=\frac{1}{1+\exp\left[-\lambda(x- \varepsilon)\right]}. \label{Eq_logistic}
\end{equation}
The fitting strategy adopted to reproduce the main features of the isosurface consists in using several sigmoids to describe the transitions between known CSS expressions as limiting cases. It is described in detail in \ref{appendix_fitting_cube}. The outcome of the fitting procedure is (for rotation energies $E_A$ and $E_{B}$ in $\si{eV}$):
\begin{align} 
&k^2_{\text{fit}}(C_{A}, C_{B}, D_{A}, D_{B}, E_{A}, E_{B}) = \left(\frac{D_A}{D_B}\right)^{\delta(E_A,E_B)} \\ \notag
&\times \left\{\left(1-\sigma_{1}(E_{A})\right)k^2_{3D-3D} +\sigma_{1}(E_{A}) \right. \\ \notag
&\times \left. \left[ \left(1-\sigma_{1}(E_{B})\right)k^2_{3D-3D} + \sigma_{1}(E_{B})k^2_{1D-1D} \right]\right\} \label{semiEmpirical}
\end{align}
with
\begin{align} 
&\delta(E_A,E_B)=\sigma_2(E_{A}) \left(-\frac{1}{2}+\frac{1}{6} \sigma_2(E_{B})\right), \\ 
&\sigma(\lambda, \varepsilon, x)=\frac{1}{1+\exp\left[-\lambda(x- \varepsilon)\right]}, \\
&\sigma_{1}(x)=\sigma(\lambda=8, \varepsilon=0.2,x), \\ 
&\sigma_{2}(x)=\sigma(\lambda=10, \varepsilon=0.5,x), \\ 
&k^2_{3D-3D} = 4 \pi R  C_{B}, \\
&k^2_{1D-1D} =   \frac{2 \pi^2 R }{-\ln\left(\pi^{2}(C_{A}+C_{B})R^{3}/2\right)} C_{B}. \label{isotropic}
\end{align}
In this expression, the $A$ label (and thus the $E_A$ assignment) must correspond to the most mobile reaction partner (so that ${D_A}\ge{D_B}$). The $k^2_{\text{fit}}$ expression has to be multiplied by $C_{A} (D_{A}+D_{B})$ to provide the absorption rate to be used in RECD. Note that in this expression, additional simplifications have been made compared to the generic form Eq. \ref{generalKappa}: $k^2_{1D-3D}$ and $k^2_{3D-1D}$ have been simply replaced by $k^2_{3D-3D}$, with an imperceptible loss of overall accuracy. Of course, that does not mean that these three CSS expressions are equal in general (the isocontour Fig.~\ref{k2_DA-DA}-(a) for $\Delta=0$ shows their measurable, though small, difference), but rather that it was chosen to cast most of their variation with $\Delta$ into a leading term, the D-ratio power delta term (or "scaling term"). This seems to  be sufficient since this simple formula reproduces qualitatively well the complex shape of the isosurfaces, as seen comparing the panels (a) and (b) of Fig.~\ref{cube2}.

\subsection{Validation of the semi-analytical expression}
We will now validate the semi-analytical formula outside of its fitting data set. 
Due to high computational cost \cite{Note5} of well-converged CSSs for high rotation energies and low concentrations, only a limited set of concentration pairs could be used. However, the formula was tested both for equal and different concentrations cases. The results are presented in table \ref{validationSet}.

\begin{table} 
\begin{tabular}{|c|c|c|c|c|}
\hline
$(C_{A},C_{B}) $             & $\mathcal{C}$      & $\varsigma$         & $M$                 & $\frac{\max_{i \in [1,n]}({k^2_{\text{eff}}})}{\min_{i \in [1,n]}({k^2_{\text{eff}}})}$ \\
$(\si{cm^{-3}})$             & Eq.\ref{correlation} & Eq.\ref{stddev-log}&  Eq. \ref{maxes} &     \\
                             &                    &                     &                      &                        \\
\hline
$(\num{1e17},\num{1e17})$    &  $0.98$            &  $0.087$            &   $2.37$               &   $49$ \\
\hline
$(\num{5e16},\num{5e16})$    &  $0.97$            &  $0.080$            &   $2.38$               &   $21$  \\
\hline
$(\num{5e17},\num{5e17})$    &  $0.98$            &  $0.078$            &   $2.36$               &   $16$  \\
\hline
$(\num{5e15},\num{5e15})$    &  $0.97$            &  $0.095$            &   $2.37$               &   $29$ \\
\hline
$(\num{2e15},\num{2e15})$    &  $0.97$            &  $0.105$            &   $2.37$               &   $31$  \\
\hline
$(\num{2e15},\num{1e16})$    &  $0.97$            &  $0.097$            &   $2.39$               &   $25$  \\
\hline
$(\num{5e16},\num{2e15})$    &  $0.98$            &  $0.091$            &   $2.35$               &   $19$ \\
\hline

\end{tabular}
\caption{Correlation coefficients $\mathcal{C}$ for $\log_{10}(k^2_{\text{eff}})$ versus $\log_{10}(k^2_{\text{fit}})$ and square root of the "mean squared error" of the model $\varsigma$ Eq. \ref{stddev-log}. The next column is the maximum discrepancy, $M=\max_{i \in [1,n]}({k^2_{\text{eff}}/k^2_{\text{fit}}})$ over the validation data sets. For comparison, the largest ratio of effective CSSs in the set, $\max_{i \in [1,n]}({k^2_{\text{eff}}})/\min_{i \in [1,n]}({k^2_{\text{eff}}})$, is given in the last column for each data set. The total radius used is always $1\ \si{nm}$ and the box dimensions range from 300 to 4000 unit lattices depending on concentrations and $L_{\text {min}}$ (technicalities on box size selections for convergence can be found in \ref{firstCriterion}).\label{validationSet}}
\end{table}

Let us consider the correlation coefficient $\mathcal{C}(x,y) =\sigma_{xy}/(\sigma_{x} \sigma_{y})$, where $\sigma_{xy}$ is the covariance and $\sigma_{x}$,$\sigma_{y}$ are standard deviations. Working with $x$ and $y$ as the logarithms of $k^2_{\text{eff}}$ and $k^2_{\text{fit}}$, respectively, is important to have a goodness of fit measure that puts on equal footing small and large CSSs. Without the logarithms, the correlation coefficient could be high only because high CSSs are well reproduced by the fit, even in cases where low CSSs are poorly reproduced. Using the logarithmic representation of the data for statistical indicators is a matter of choice, and this choice actually corresponds to focusing on reproducing the orders of magnitude of effective CSSs over the whole range of parameters rather than aiming at extreme precisions for the highest ones. Also, this is consistent with the fitting procedures which were also done on logarithms for the same goals. 

In table \ref{validationSet}, we note that the correlation coefficient for $\log_{10}(k^2_{\text{eff}})$ versus $\log_{10}(k^2_{\text{fit}})$ 
\begin{equation} \label{correlation}
\mathcal{C} = Corr(\{\log_{10}(k^2_{\text{eff},i})\}, \{\log_{10}(k^2_{\text{fit},i})\}),
\end{equation}
(where $\{\log_{10}(k^2_{\text{eff},i})\}$ and $\{\log_{10}(k^2_{\text{fit},i})\}, {i \in [1,n]}$ are the sets of the $n=1728$ logarithms of effective and fitted CSS values, respectively) are very close to one. This shows that, at least in terms of orders of magnitude, the formula captures almost perfectly the CSS evolution over the parameter ranges. 

\begin{figure}
\includegraphics[width=0.5\textwidth]{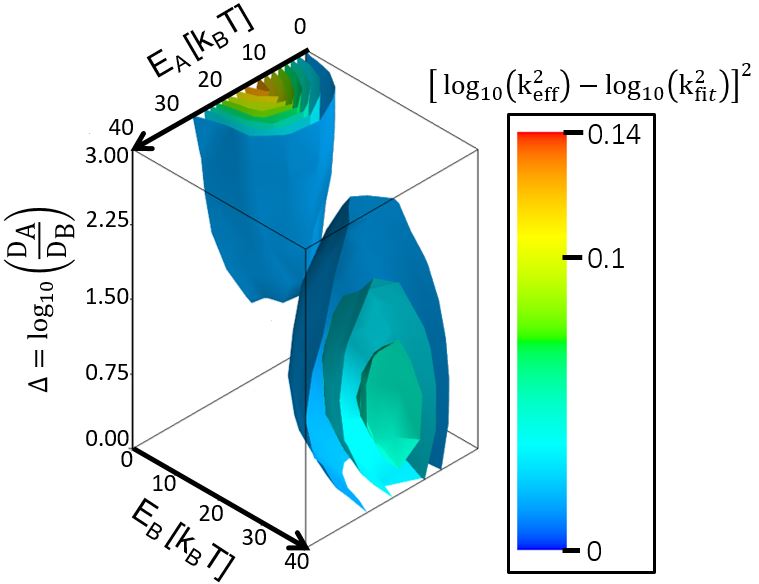}
\caption{Isosurfaces of residuals of logarithms from Eq. \ref{logarithmic_residue} for the reference conditions ($T=573\ K, C_{A}=C_{B}=\num{1e17} \si{cm^{-3}}, R_{A}=R_{B}=1 \si{nm}$)}
\label{residue_isosurfaces}
\end{figure}

Another measure of the goodness of fit can be obtained considering the square root of the "mean squared error" of the model \cite{Sammut}:
\begin{equation} \label{stddev-log}
\varsigma = \left(\frac{1}{n}\sum_{i=1}^n\left(\log_{10}(k^2_{\text{fit},i})-\log_{10}(k^2_{\text{eff},i})\right)^2\right)^{1/2},
\end{equation}
where $i$ iterates through the $n=1728$ conditions of the set. From table \ref{validationSet}, we see that the square roots of the mean squared errors are similar ($0.09$) for all validation sets. This gives an estimate of the "average error" (coefficient of variation) of the semi-analytical formula: about 20\%, as $10^{0.09}\simeq1.2$. Globally, this indicates a quite good validation of the semi-analytical fit. At first sight, the quite large local maximum discrepancies 
\begin{equation} \label{maxes}
M=\max_{i \in [1,n]}\left(\frac{k^2_{\text{eff},i}}{k^2_{\text{fit},i}},\frac{k^2_{\text{fit},i}}{k^2_{\text{eff},i}}\right),
\end{equation}
displayed in table \ref{validationSet} could be considered as a source of inaccuracy. To that concern first, it should be noted that these large deviations mostly correspond to intermediate values of rotation energies, as can be seen from Fig.~\ref{residue_isosurfaces} where the isosurfaces of the residuals of logarithms
\begin{equation}
r_{i}=\left[ \log_{10}(k^2_{\text{eff},i})-\log_{10}(k^2_{\text{fit},i}) \right]^2, \label{logarithmic_residue}
\end{equation}
are represented. In the figure, they are found to be everywhere very close to $0$ (blank zones are below $0.00746$, whose square root elevated to power ten corresponds to values below the 20\% average discrepancy) except when $E_{A}\simeq 15\ k_{B}T$,  $E_{B}\simeq 0$ and $\Delta \gtrsim 2$. In that very small orange zone, it reaches $0.14$ whose square root corresponds to the logarithm of the maximum discrepancy $M$ for the first condition in table \ref{validationSet} ($\log_{10}(2.37)\simeq\sqrt{0.14}$): the zones of quite high deviation of the fit are always quite small and actually correspond to transition zones.

The small extent of significant residue zones indicates that the approach consisting in entering limiting cases in the fitting formula is overall relevant and that it is mostly the choice of the transition function involving two basic logistic functions that could be improved. It appears that parity of the logistic function (Eq.\ref{Eq_logistic} is an odd function of $x$ when changing its origin to $(\varepsilon,1/2)$) does not allow to match perfectly the evolution of the effective CSS for both $x$ smaller than $\varepsilon$ and $x$ larger than it. This asymmetry needed to reproduce finer details of the effective CSS over the whole range may arise from the need of $k_{1DR-0}$ expressions to describe them for example near $E_{A}\simeq 15 k_{B}T$,  $E_{B}\simeq 0$ at large $\Delta$. Nevertheless, to keep it as simple as possible, these limiting cases where not included in the semi-analytical formula.
A possibility that would not require to include more limiting cases would be to use other functions to model the transition (asymmetric sigmoid like the generalised logistic function) that do not impose this symmetry. This was tested and achieved some partial improvement but at the cost of doubling the number of parameters. To that concern, fitting the transition zones with splines would probably be an even better option, but the number of parameters would be as large as a multiple of the number of fitting points.

Notwithstanding these possible refinements of the fit in the transition regions, casting all possible CSSs into a general formula should be seen as a significant improvement of the CSS description compared to over-simplifiying assumptions consisting in approximating them with the $3D$ CSSs or oppositely with $1D-0$ ones. Indeed, we can see from table \ref{validationSet} that varying rotation energies couples and diffusion ratio the CSSs vary with a factor of 50 as the dimensions of mobilities go from $1D-1D$ to $3D-3D$. Thus, the approximation of treating them as purely 3D would result for in overestimating also up to a factor 50, in the present example, and even more for $\Delta>3$. This as to be compared with the 20\% overall accuracies allowed by the semi-analytical formula. 
Depending on the application, such precision might be sufficient or not. For example for a very precise computation of loops nucleation out of irradiation, all the errors on the estimated absorption rates for monomers up to critical size clusters will cumulate and may have an impact on the overall estimated nucleation rates. On the other hand, for microstructural evolution under irradiation, where the cascade cluster production may allow clusters to grow bypassing the classical nucleation path and where, due to colossal supersaturations, the expected critical radius may be extremely small, then reproducing the order of magnitude of CSSs may be a reasonable approximation, considering all others sources of both modeling and experimental uncertainties.

It is also interesting to note that quite counter-intuitively, simple reductions of the dimensionality of the mobilities are often misleading whereas dimensionality equivalences are relevant:
having both species 1D-migrating and $D_{B}$ as small as $D_{B}/D_{A}=10^{-3}$ does not allow to treat $B$ as immobile, as the effective CSS is far from the analytical $1D-0$ CSS, whereas it matches well the $1D-1D \Leftrightarrow 2D-0$ CSS with proper accounting of the scaling term. 
This important point will be further investigated in the next section.

Finally, note that this semi-analytical formula is directly valid only around $T=573 {\si K}$. As such, it is also bound to the first nearest neighbor jump distance used in the OKMC simulations for the fitting set. To extend it to other temperatures and other lattices than BCC, a simple modification is proposed in \ref{extension_to_other_temperatures}.

\subsection{Effective exponents analysis: the reaction dimensionality diagram}
\begin{figure}
\includegraphics[width=0.5\textwidth]{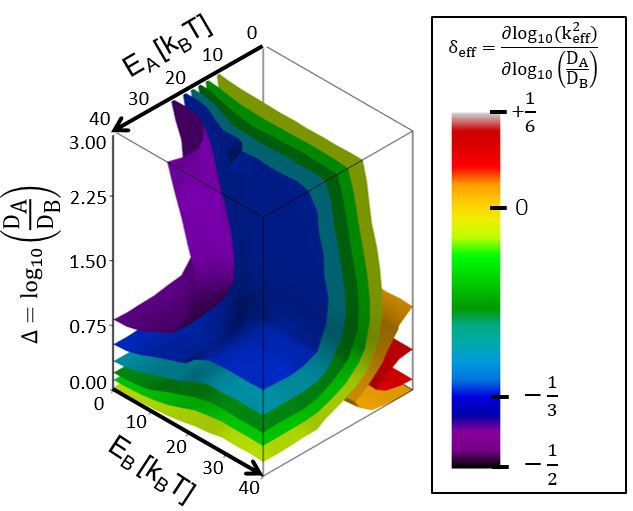}
\caption{Effective exponents $\delta_{\text{eff}}$ isosurfaces, allowing to identify CSS domains.}
\label{delta-eff}
\end{figure}

We now propose a more practical method to physically interpret the variation of the CSS in the $(E_{A}, E_{B}, \Delta)$-space. It consists in using the logarithmic derivative of the CSS to estimate an effective exponent $\delta$ as a function of the considered $(E_{A}, E_{B}, \Delta)$ point. This stems from the properties of the logarithmic derivative of a power-law $a x^b$: $\frac{d log\left(a x^b\right)}{dlog(x)}=b$. 
The figure \ref{delta-eff} represents the actual map of effective $\delta$ values, 
\begin{equation}
\delta_{\text{eff}}\left(E_A,E_B,\Delta \right) = {\frac{\partial \log_{10} \left( k^2_{\text{eff}}\right)}{\partial \log_{10} \left(  \frac{D_A}{D_B}\right)}}, \label{logarithmic_derivate}
\end{equation}
from the set of estimated CSSs. 

The connection with the limiting cases is quite straightforward (actually, it served as a backbone for the fitting function in \ref{appendix_fitting_cube}):
\begin{itemize}
\item $(E_{A}=0,\ E_{B}=0) \ $ corresponds to $3D-3D$ absorption rates which simply depend on $(D_{A}+D_{B})$, that makes $(1+(D_{A}/D_{B})^{-1})$ when factoring for $D_B$ the CSS, which is close to $1=(D_{A}/D_{B})^0$ when $D_{A} \gg D_{B}$ in other words a close to zero $\delta$ value,
\item $(E_{A}\gg k_{B}T,\ E_{B}\gg k_{B}T)$ corresponds to $1D-1D$ absorption rates with a characteristic exponent of $-1/3$, consistently with the related limiting case \cite{Adjanor1},
\item $(E_{A}\gg k_{B}T,\ E_{B}=0)$ similarly corresponds to $1D-3D$ with $D_{A}>D_{B}$ absorption rates leading to a $-1/2$ exponent \ref{section_Sink_strengths_in_the_3D_anisotropic_case}.
\end{itemize}

The figure may also be seen as a "reaction dimensionality diagram", as it permits, basing on the values of the exponent, to identify the extent of zones in the $(E_{A}, E_{B}, \Delta)$-space associated with the $1D-1D$ reactions (the zone delimited by the blue surface in the figure) or  with the $1D-3D$ ones (the purple zone). The presence of this $1D-3D$ domain, is due to the fact that, increasing $\Delta$, $D_{B}$ gets lower than $D_{A}$, so the influence of $B$'s mobility characteristics diminishes, but if $E_{A}$ is large ($A$ is then a 1D-diffuser), small values of $E_{B}$ can still make $B$ an efficient 3D-diffuser. In other word, the diagramm reflects the fact that the dimensionality of mobilities (i.e. rotation energies) can compensate a decreasing diffusion coefficient ratio and impose specific kinetics. The yellow envelop surface in the diagramm can be seen as a quasi-$3D-3D$ zone, and the rest of the surfaces should be considered as transition zones, apart from a small zone where $D_{A} \simeq D_{B}$, $E_{A}\simeq 0$, $E_{B} \simeq 40 k_B T$ (in red in the figure) where the exponent approaches $1/6$. Due to its very small extent in the $\Delta$-dimension this red zone wasn't identified so far in the analysis of limiting cases, but it can be interpreted considering the analog case highlighted in \ref{section_Sink_strengths_in_the_3D_anisotropic_case}:
\begin{flalign}\label{anisotropicCSS}
&\frac{\partial C_{A}}{\partial t}= - 8 R {\overline D} \left( \frac{D_z} {D_{\rho}} \right)^{1/6} C_{A} C_{B}, \text{for $D_z \gg D_{\rho}$},
\end{flalign}
with ${\overline D}=(D_{\rho}^2 D_{z})^{1/3}$. According to the authors of this development \cite{GoseleSeeger,Woo2}, it should be rightful for $D_z \gg D_{\rho}$. Using the analogy approach of the companion paper, we may take $D_z=D_A+D_B$ and $D_{\rho}=D_{B}$. Actually, the red zone is for quite small $\Delta$ values so, if we admit that the relation also holds for moderate diffusion "anisotropy" ($D_z \gtrsim D_{\rho}$, ${\overline D}$ should vary much then) then, the characteristic exponent $+1/6$ is explained. The fact that this zone actually has a very limited range and does not extend to larger $\Delta$ values is just due to the chosen representation: because the CSSs are studied with $(D_A+D_B)$ constant, $D_B$ vanishes along the $z$-axis and the evolution is then only driven by $E_A \simeq 0$ and its $\delta=0$ typical exponent.

\subsection{Closure of the CSS domains} \label{Closure}
\begin{figure}
\includegraphics[width=0.5\textwidth]{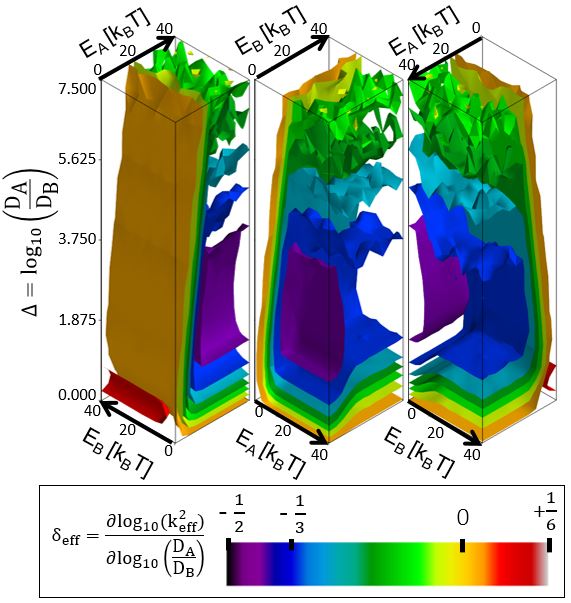}
\caption{Three views of the effective exponents' isosurfaces, $\delta_{\text{eff}}$.}
\label{montage_BIGGER}
\end{figure}

In the previous section, we noted that even for diffusion coefficients ratios as low as ${\mathcal D}=10^{-3}$, the least mobile specie may not be treated as immobile, as the most relevant analytical CSS is the properly corrected $1D-1D$ expression. It is important for modeling concerns to investigate to which extends this holds, and what is the typical diffusion ratio where the $1D-0$ CSS starts to be more relevant.
The semi-analytical formula that we have established allows for a reasonably accurate estimation of CSSs depending on $(E_A, E_B, \Delta)$, but due to its numerical cost, the fitting set was limited to $0 \le \Delta \le 3$.
On that part of the $(E_A, E_B, \Delta)$-space, the two main absorption-rate domains are clearly visible ($3D-1D$ and $1D-1D$) but none of them are "closed" at $\Delta=3$. As a consequence, the validity of the formula as such is restricted to that range, because there must be some D-ratio below which $B$ should be considered as immobile and CSSs tend to $3D-0$ and $1D-0$ expressions respectively. For the former, the transition does not need to be explicitly accounted for as $3D-3D$ CSS expressions naturally encompass the case $D_{B}=0$, whereas for the $1D-1D$ to $1D-0 $ transition the reaction rates do not even have the same reaction order. Thus the range of $\Delta$ was extended to $\Delta=7.5$ in a new calculation set where the sampling criteria were slightly relaxed. As displayed in Fig.~\ref{montage_BIGGER} the domain for the $1D-3D$ CSS now closes by $\Delta=4.5$ and that of the $1D-1D$ CSS closes about $5$. From $\log_{10}(D_{A}/D_{B})=3.75$ to $7.5$, some isosurfaces are wavy because the sampling set is smaller than for the former conditions. This could not be easily improved as the present result already required about 2.2 million CPU hours on Xeon Sandy Bridge 2.6 GHz cores. Being more precise on the closures in the $(E_A, E_B, \Delta)$-space would require quite a lot more computational resources, so we should rather focus computational efforts on more restricted case, the $(E_{A}=40 k_{B}T, E_{B}=40 k_{B}T)$ case.

\begin{figure}
\includegraphics[width=0.5\textwidth]{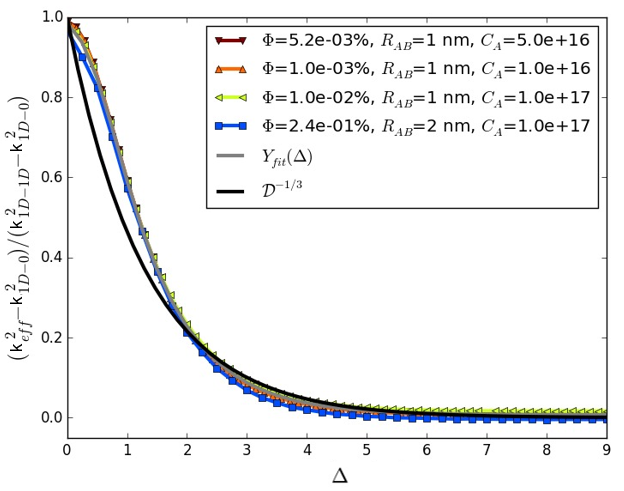}
\caption{The transition of the normalised effective CSSs from $1D-1D$ to $1D-0$ for four different conditions. The generalised sigmoid fit follows Eq. \ref{generalizedSigmoid}. In the caption, $\Phi$ represents the clusters volume fractions and the concentrations $C_{A}=C_{B}$ are given in $\si{cm^{-3}}$. 
}
\label{k2_calc_S_k2_2Dand1D_vs_D1_-S-_D2_FEW_CURVES}
\end{figure}

To that end, a few sets of simulations (for a few $(C, R)$ couples, $C=C_{A}=C_{B}$) for $\Delta$ ranging from $0$ to $9$ were run. The results are shown in Fig.~\ref{k2_calc_S_k2_2Dand1D_vs_D1_-S-_D2_FEW_CURVES} which represents the normalised quantity:
\begin{equation}\label{conformal}
Y(\Delta)=\frac{k^2_{\text{eff}}-k^2_{1D-0}}{k^2_{1D-1D}-k^2_{1D-0}}. 
\end{equation}
This quantity is equal to one when $k^2_{\text{eff}}=k^2_{1D-1D}$ and when it tends to zero then $k^2_{\text{eff}}$ tends to  $k^2_{1D-0}$. When varying concentrations and radii, this representation helps to check if a common trend for the transition towards ${1D-0}$ stands out. We see that it is indeed the case for the typical volume fraction conditions investigated. The question of whether this simple behavior extends to a much wider range of $(C_A, C_B, R)$ conditions could be delicate and is not addressed here. We rather focus on proposing a practical correction for the vanishing scaling factor $({\cal D})^{-1/3}$ that would clearly lead to an underestimation of CSSs from some point when $\Delta$ is large. For the investigated conditions, the $1D-1D$ to $1D-0$ transition happens to follow a common trend that is well fitted by the generalised sigmoid,
\begin{equation}\label{generalizedSigmoid}
Y_{\text{fit}}(\Delta)=a-\frac{b}{(c+d \exp(e \Delta-f))^g} 
\end{equation}
with $a=0.0071, b=-1.26, c=2.23, d=0.932, e=3.81, f=0.98, g=0.26$ as shown in Fig.~\ref{k2_calc_S_k2_2Dand1D_vs_D1_-S-_D2_FEW_CURVES}.

\begin{figure}
\includegraphics[width=0.5\textwidth]{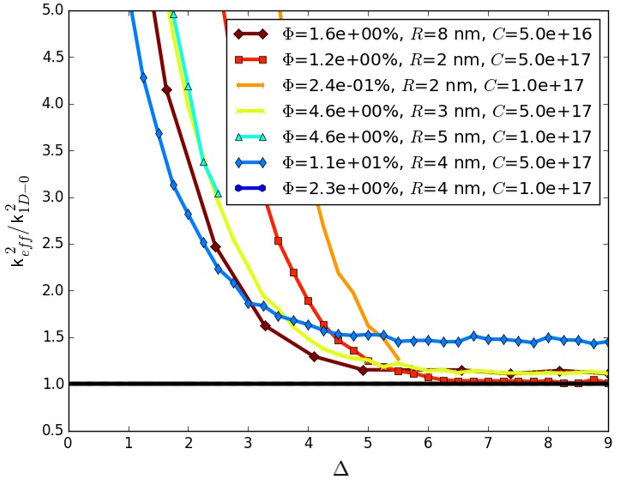}
\caption{Convergence of the ratio of the effective CSS over the analytical $1D-0$ CSS for different conditions. In the legend, $\Phi$ represents the volume fractions and the concentrations are given in $\si{cm^{-3}}$.}
\label{k2_calc_S_k2_2Dand1D_vs_D1_-S-_D2_FEW_CURVES_ZOOM}
\end{figure}

From the representation of $k^2_{\text{eff}}/k^2_{1D-0}$ in Fig.~\ref{k2_calc_S_k2_2Dand1D_vs_D1_-S-_D2_FEW_CURVES_ZOOM}, we can conclude that, depending on the $C$ and $R$, the effective CSS reaches the analytical $k^2_{1D-0}$ within a typical 5-10 \% uncertainty of estimates around $5<\Delta<6$, except for the blue curve whose atypical behavior is explained by a very large volume fraction. Formally, we can identify a critical $\Delta$ value by equating the $1D-0$ and $1D-1D$ CSSs which leads to 
\begin{equation}
\Delta^{*} = -3 \log_{10} \left( \frac{3}{\overline{\alpha}} R^3 C\right), \label{critical}
\end{equation}
where ${\overline{\alpha}}\simeq\frac{4}{\ln(\pi^2 C R^3/{2})}$ is the usual parameter entering in the $1D-1D$ CSS expression \cite{Adjanor1}.
The evaluation of $\Delta^{*}$ for the conditions of Fig.~\ref{k2_calc_S_k2_2Dand1D_vs_D1_-S-_D2_FEW_CURVES_ZOOM} leads to values mostly between $6$ and $8$, which corresponds reasonably well to the critical values from the figure. Nevertheless, the approach from the Eq.\ref{critical} and the actual evolution towards the critical values do not compare further: it is clear that the transition is very gradual and that both $1D-1D$ and $1D-0$ mechanisms operate simultaneously even above the critical value. In the simplest way, the cross-over between mobile and fixed sinks related CSSs may be cast with semi-analytical formula as:
\begin{eqnarray}
Max\left(k^2_{\text{fit}}(C_A, C_B, R, E_A, E_B, D_A, D_B),\right.\\ 
\left.k^2_{1DR-0}(C_A, C_B, R, E_A, D_A)\right). \label{Eq_transition_Max}
\end{eqnarray}
A more elaborated way of reproducing the transition would be to use the fit Eq. \ref{generalizedSigmoid} and replace it in Eq. \ref{conformal}:
\begin{eqnarray}
k^2_{\text{fit}} = Y_{\text{fit}}(\Delta)(k^2_{1D-1D}-k^2_{1D-0})+k^2_{1D-0}, \label{Eq_transition_fit}
\end{eqnarray}
but this should be valid only for pure $1D-1D$ and establishing a $Y_{\text{fit}}$ function valid for any rotation energy couple would require a more complex function with much more parameters.

The notion of mean-free-path between rotations $\ell_{ch}$ is often used to interpret the transition in the reaction order of the $1DR$ CSS. This quantity is most meaningful when compared with the mean distance between sinks (inverse square-root of the relevant sink-strength): when $\ell_{ch}$ is much smaller than the mean distance between sinks, the mixed mobility gets close to a pure 3D-random walk. Conversely, if $\ell_{ch}$ is much longer than the mean distance between sinks, the mixed mobility diffusion boils down to a pure 1D-diffusion process. Thus, as it becomes manifest when considering the "master-curve" \cite{Heinisch,Barashev,Trinkaus2002} for $1DR-0$ CSS, the effective reaction order of the kinetics depends, in general, on the total sink-strength and its time evolution. When the interaction of two $1DR$ diffusers is considered, it becomes difficult to generalise these considerations. Fortunately, as show in the companion paper, when two 1D-diffusers are considered, the total sink-strength effects are expected to be quite weak, just as in well-known case where only 3D-diffusers are considered. Thus, the proposed semi-analytical formula established with CSS calculations on pairs of diffusers is expected to be valid in the multiple sinks case, provided that $\Delta$ is not too large. Figure \ref{k2_calc_S_k2_2Dand1D_vs_D1_-S-_D2_FEW_CURVES} shows the slow transition between $1D-1D$ and kinetics $1D-0$ which are, respectively, second-order with weak multi-sink effects and third-order with strong multi-sink effects. Equations \ref{Eq_transition_Max} and \ref{Eq_transition_fit} could be seen as workarounds for the transitions between both reaction orders and multi-sink characters, but this would need a dedicated validation.

\section{Application to cluster dynamics} \label{sectionCD}
We now briefly expose the results of the application of the semi-analytical CSS on cluster dynamics simulations. This development allows to properly account for the agglomeration of defect clusters with mixed 1D/3D mobility i.e. any couple of rotation energies. Some additional technical details on the implementation in the RECD code CRESCENDO \cite{Jourdan} can be found in the corresponding section of the companion paper. Here we focus on the ability of the present extension to reproduce an OKMC microstructure simulation close to the state-of-art in terms of complexity of random walks. A concentration of $\num{2e16}\ \si{cm^3}$ interstitial monomers is initially placed in a quasi-cubic box of about $2000$ lattice constants ($a_{0}$) length. A test parameterisation was chosen to allow the formed clusters to be mobile up to the size $60$. Its mobility parameters are summarised in table~\ref{tableParamSummary}. Increasing cluster content ("cluster size") from $1$ to $60$ monomers, the diffusion coefficients are decreasing according to the power-law $n^{2/3}$ from $\num{1.793e-5}$ to $\num{1.168e-6}\ \si{cm^{2} s^{-1}}$ and their rotation energies increase from $0$ to $2\ \si{eV}$ (linearly up to size $12$ and then the energy reaches a constant value of $2\ \si{eV}$). Regarding rotation energies, these parameters are intended to sketch some state-of-the-art OKMC parameterisations such as the parameterisation in table~\ref{tableMalerba}. Notably, the increase of $0.2\ eV$ per monomer adopted here was originally assumed by Wen et al. \cite{Wen} (who also reported alternative assumptions of $0.05\ eV$ increase per monomer).

\begin{table}
\begin{tabular}{|c|c|c|c|}
\hline
$n$            & $R$       &$D$                  & $E$             \\
$ $            &$(\si{nm})$     &$(\si{cm^{2} s^{-1}})$     &  $(\si{eV})$      \\
\hline
$1 $           & $0.516$   &$\num{1.793e-05}$      &  $0.0$          \\
$2 $           & $0.567$   &$\num{1.129e-05}$      &  $0.1$          \\
$3 $           & $0.606$   &$\num{8.616e-06}$      &  $0.2$          \\
$4 $           & $0.639$   &$\num{7.112e-06}$      &  $0.3$          \\
$5 $           & $0.668$   &$\num{6.128e-06}$      &  $0.4$           \\
$6 $           & $0.682$   &$\num{5.426e-06}$      &  $0.5$           \\
$7 $           & $0.705$   &$\num{4.896e-06}$      &  $0.6$           \\
$8 $           & $0.725$   &$\num{4.479e-06}$      &  $0.7$           \\
$9 $           & $0.745$   &$\num{4.140e-06}$      &  $0.8$           \\
$10$           & $0.764$   &$\num{3.859e-06}$      &  $0.9$           \\
$11$           & $0.781$   &$\num{3.622e-06}$      &  $1.0$           \\
$12$           & $0.798$   &$\num{3.417e-06}$      &  $2.0$           \\
$\vdots$       & $\vdots$  &$\vdots$               &  $2.0$           \\
$60$           & $1.28$    &$\num{1.168e-6}$       &  $2.0$           \\

\hline                                                                     
\end{tabular}                                                              
\caption{Summary of the parameterisation of cluster mobility. $D$ is the diffusion coefficient and $E$ the rotation energy of a cluster containing $n$ monomers. Note that, following \cite{chiapetto2015nanostructure}, the radii $R$ are given assuming clusters are spherical, but they are augmented by capture efficiencies and spontaneous recombination distances. }\label{tableParamSummary}  
\end{table}

\begin{figure}
\includegraphics[width=0.5\textwidth]{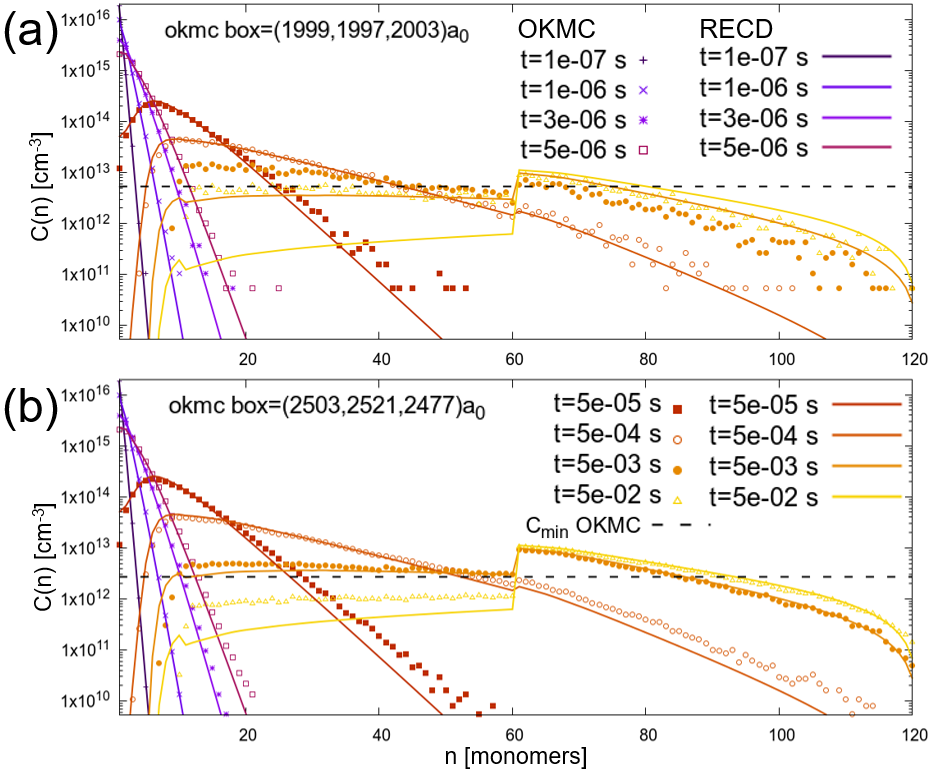}
\caption{Time evolution of the defect clusters distributions. Starting from a population of $2 \times 10^{16}\ \si{cm^{-3}}$ SIAs, the time evolution of the defect clusters population with mixed mobility was obtained by averaging 1000 OKMC runs (points marked by symbols). The comparison with the RECD implementation of $k^2_{\text{fit}}$ is given by the continuous lines. The dashed horizontal line represents the minimum concentration $C_{\text {min}}$ (the inverse of the box volume) that can be considered by an individual OKMC run. Top: 1000 OKMC runs with a "quasi-cubic" box size of around 2000 lattice units. Bottom: 1000 OKMC runs with boxes dimensions around 2500 lattice units.}
\label{2e16_redogaia_b2000}
\end{figure}

To be comparable with RECD, OKMC simulation must consist of hundreds to thousands of runs with different random seeds: an individual OKMC run generally ends up with a few clusters resulting in a sparse distribution, whereas in RECD significant concentrations of clusters are often found in quite smeared and continuous distributions. The figure \ref{2e16_redogaia_b2000}-(a) shows the comparison between the two simulation methods. It is worth mentioning that, although the set of 1000 lengthy OKMC runs with box size around $2000 a_{0}$ (top panel (a)) represents a considerable amount of computer resources compared to the RECD simulation (OKMC requires here several million times more individual CPU time than RECD), it is still not enough to fully characterize the distribution as it appears discontinuous due to the lack of sampling at the greatest sizes. More precisely, at intermediate times (around $t=\num{5e-3}\si{s}$ and $\num{5e-2}\si{s}$), the OKMC distribution contains many concentrations $\{C(n), n \in [40,60]\}$, which are below the minimum concentration that can be modeled in an individual OKMC run ($C_{\text {min}}=1/V$, the dashed horizontal line of Fig.~\ref{2e16_redogaia_b2000}-(a) and -(b)). Artificially, in individual runs, there is a too sparse population of these clusters which prevents them from correctly interacting with the others. This artifact results in abnormally slow kinetics for the products of these reactions and explains the mismatch with RECD calculations at intermediate times in panel (a). Indeed, if the box size is increased to around $2500 a_{0}$ (panel (b) from Fig.~\ref{2e16_redogaia_b2000}), then there are enough of these mobile clusters at all steps where mobile-mobile interactions are dominant: we can see that $C_{\text {min}}$ is decreased compared to panel (a) and that the OKMC points are now above this threshold at $t=\num {5e-3}\si{s}$. Consequently, the agreement with RECD is even further improved, owing to the decreased $C_{\text {min}}$ value. 

\begin{figure}
\includegraphics[width=0.5\textwidth]{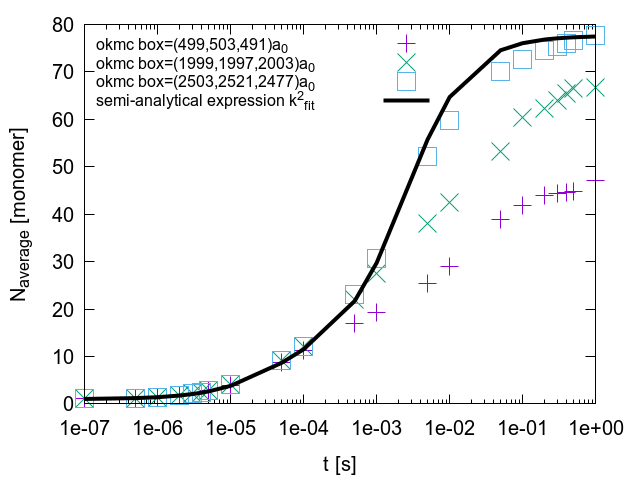}
\caption{Time evolution of the average size associated with the cluster distributions. Considering the differences in the convergence of OKMC simulation with increasing boxes sizes, the necessity to consider large enough boxes is manifest. Each set is averaged on 1000 runs.}
\label{Nave_f_t_cres_test_ARTICLE_convergence_OKMC}
\end{figure}

In terms of average cluster size $N_{\text{average}}$: 
\begin{equation}
N_{\text{average}}=\frac{\int n C(n) dn}{\int C(n) dn}, 
\end{equation}
the RECD/OKMC comparison (Fig.~\ref{Nave_f_t_cres_test_ARTICLE_convergence_OKMC}) appears as almost completely satisfactory. Also, the importance of considering large enough box sizes for this type of OKMC simulation with an advanced mobility parameterisation is even more manifest (symbols of Fig.~\ref{Nave_f_t_cres_test_ARTICLE_convergence_OKMC}). We clearly see that using too small boxes results in wrong plateau values of $N_{\text{average}}$. Box size effects are commonly considered as a source of artifacts due to the limited distance between periodic boundary conditions, but here the box size is much larger than the average cluster radius, so it is rather a \textit{box volume} effect in connection with the $C_{\text {min}}$ criterion.

\subsection{On the absence of manifest multi-sink effects for $1DR-1DR$ reactions}
In the application section of the companion paper, the good agreement between RECD and OKMC was put forward as an argument for the absence of significant "multi-sink" terms, the effect of the rest of the cluster's population on the rate of individual reactions for a given cluster reaction pair. Nevertheless, because the distributions were quite peaked in that pure $1D-1D$ parameterisation, one could argue that it was not an ideal test for potential multi-sink effects. Here, we are much closer to such an ideal test situation for multi-sink effects: the distribution is very flat at the intermediate time steps (around $t=\num {5e-3}\si{s}$), so the interference of the whole population of clusters on pair reactions should be manifest, if it happens to be actually significant. Here, we confirm that no multi-sink terms were necessary in $k^2_{\text{fit}}$ to accurately reproduce the OKMC distribution, so this is an additional argument to consider that they should be negligible in low to moderate volume fraction conditions.
In the companion paper, the argumentation also relied on the direct assessments of the very weak perturbation of these bimolecular absorption rates by a third specie. The success of this test is not surprising from an analytical point of view. Indeed, it can be shown (\ref{homogeneity}) that, because the "degree 1 homogeneity" property (formally, this writes: $k^2(C,\lambda C) \simeq \lambda k^2(C, C)$) is reasonably well verified for all the components of $k^2_{\text{fit}}$, splitting a class of interacting clusters into arbitrary subclasses results in partial sink-strengths whose sum is equal to the total sink-strength (i.e the sink-strength without splitting). Thus, there is no analytical inconsistency between the expression $k^2_{\text{fit}}$ and the assumption of the absence of multi-sink terms at the moderate volume fractions of our interest.

\subsection{Comparison with other approaches} \label{comparison}
\begin{figure}
\includegraphics[width=0.5\textwidth]{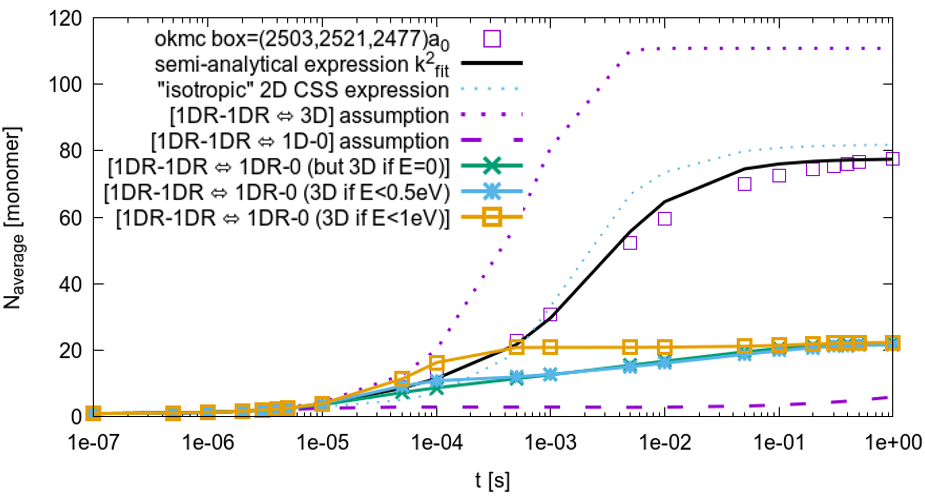}
\caption{Time evolution of the average size associated with the cluster distributions. Regarding RECD calculations, the results of the new CSS propositions (the $1D-1D \Leftrightarrow 2D-0$ "isotropic" expression from the companion paper \cite{Adjanor1} and the semi-analytical $k^2_{\text{fit}}$) are compared to the implementation of various CSS approximations from the literature: $\left[1DR-1DR \Leftrightarrow 3D-0\right]$, $\left[1DR-1DR \Leftrightarrow 1D-0\right]$, and the $\left[1DR-1DR \Leftrightarrow 1DR-0\right]$ assumption starting from different rotation energies thresholds.}
\label{Nave_f_t_cres_test_ARTICLE_compare_1DR-0_assumptions}
\end{figure}

Finally, to highlight the impact of the approximations made to model CSSs for mutually mobile species, the implementation of $k^2_{\text{fit}}$ is compared to the $\left[1DR-1DR \Leftrightarrow 3D-0\right]$ and the $\left[1DR-1DR \Leftrightarrow 1DR-0\right]$ assumptions with an equivalent $D=D_{A}+D_{B}$ as often found in the literature (the respective original arguments from the literature were already discussed in the introduction of the companion paper \cite{Adjanor1}). Consistently with the order of magnitude considerations developed in the previous paper, the $3D-0$ approach is found to overestimate the plateau value of the $N_{\text{average}}$ and also the kinetics to reach this plateau by more than a factor of ten, as seen in Fig.~\ref{Nave_f_t_cres_test_ARTICLE_compare_1DR-0_assumptions}. 

Oppositely, the $\left[1DR-1DR \Leftrightarrow 1D-0\right]$ assumption is not expected to produce significant clustering here, owing to the extremely small intensity of $1D-0$ absorption rates compared to their $3D-0$ counterparts. A more elaborated assumption could be the $\left[1DR-1DR \Leftrightarrow 1DR-0\right]$ one. This approximation could be done in several ways. The current parameterisation considers monomers and dimers as $3D$-diffusers, so $3D-3D$ CSSs (equivalent to $3D-0$ with the sum of diffusion coefficients) should be used for the reactions involving one of them. For reactions involving two clusters $A$ and $B$ with mixed mobility ($E_{A}>0$, $E_{B}>0$), the $\left[1DR-1DR \Leftrightarrow 1DR-0\right]$ assumption would consist in summing up the individual sink-strengths $k^2_{1DR-0}(E_A,D_A)$ and $k^2_{1DR-0}(E_B,D_B)$ \cite{Heinisch,Barashev,Trinkaus2002}, just as if a $A$-cluster was considered as mobile with respect to $B$-clusters, and reciprocally (as explicitly assumed in \cite{Dunn} among other papers). Note in passing that the multi-sink terms (here, the sum of "partial sink-strengths" \cite{Barashev,Borodin}) which are inherent to the $1D-0$ components of the $1DR-0$ CSS have been taken into account in these tests. The resulting kinetics being extremely slow (see Fig.~\ref{Nave_f_t_cres_test_ARTICLE_compare_1DR-0_assumptions}), one could also argue that this simple CSS "summing rule" should only hold for large enough rotation energies, and that interactions involving clusters with rotation energies below some threshold should considered as $3D-0$ as well. This corresponds to the $\left[1DR-1DR \Leftrightarrow 1DR-0\right]$ with threshold ($3D$ if $E<0.5\ \si{eV}$ and $E<1\ \si{eV}$) assumptions in Fig.~\ref{Nave_f_t_cres_test_ARTICLE_compare_1DR-0_assumptions}. None of these threshold values seem to help the "$1DR-0$-equivalence" to reproduce the OKMC kinetics of $N_{\text{average}}$. 

This is not surprising since the $1DR-0$ CSS expression does not even have the right order of reaction to describe neither $1D-1D$ nor $1DR-1DR$ absorption kinetics, as it was emphasized in the companion paper. This does not put into question the validity of the so-called "master-curve" approach \cite{Heinisch,Barashev,Trinkaus2002} for the $1DR-0$ kinetics which actual purpose is considering absorption towards fixed sinks, but rather shows that their direct use for general $1DR-1DR$ kinetics does not seem to pass a dedicated validation.

\begin{figure}
\includegraphics[width=0.5\textwidth]{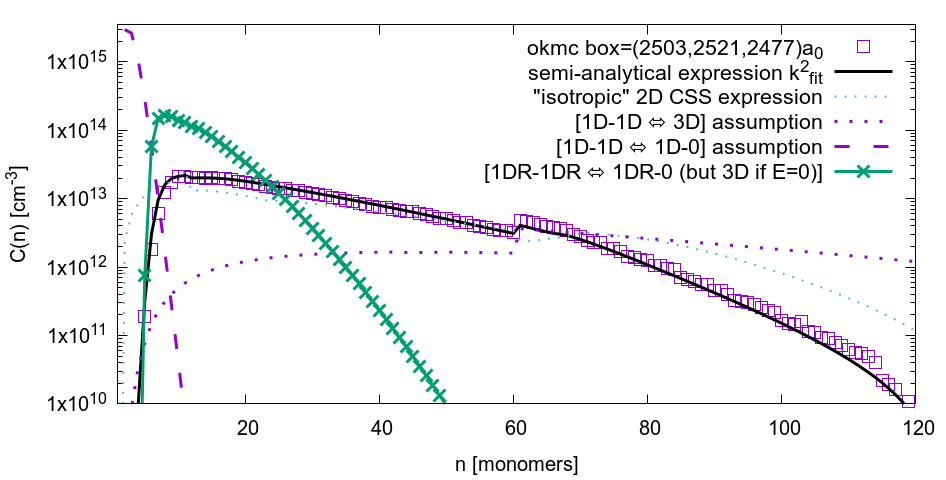}
\caption{Defect clusters distribution at $t=\num{1e-3} s$. Regarding RECD calculations, the results of the new propositions (the $1D-1D \Leftrightarrow 2D-0$ "isotropic" expression from the companion paper \cite{Adjanor1} and the semi-analytical $k^2_{\text{fit}}$) are compared to the implementation of various CSS approximations from the literature: $\left[1DR-1DR \Leftrightarrow 1DR-0\right]$ and $\left[1DR-1DR \Leftrightarrow 3D-0\right]$ that were discussed in the companion paper \cite{Adjanor1}.}
\label{comp_NEW}
\end{figure}

Another interesting point is that, for the considered parameterisation, it seems that it is primarily the $1D-1D$ reaction order that drives the overall kinetics, and then, to a lower extent, accounting for the actual $1DR-1DR$ kinetics with $k^2_{\text{fit}}$ allows to reproduce finer details of the dynamics. Indeed, as seen in Fig.~\ref{Nave_f_t_cres_test_ARTICLE_compare_1DR-0_assumptions} and Fig.~\ref{comp_NEW}, the simplest version of the $1D-1D$ equivalence to the "isotropic" $2D-0$ CSS Eq. \ref{isotropic} still catches some features of the overall evolution, although it is far from being as accurate as the semi-analytical expression. Nevertheless, it is important to stress out that, because vacancies are deliberately neglected, this application serves for validation purposes and is \textit{a priori} not representative of any irradiation condition of practical interest.

\subsection{Simulation of extreme electron irradiation} \label{irradIV}
Let us now assess the RECD implementation of semi-analytical CSS for electron irradiation modeling. Regarding interstitials and their mobile clusters, the parameterisation is given by table \ref{tableParamSummary}, while for vacancies, only monomers are considered as mobile with the following characteristics: $D=\num{1.777e-09}\ \si{cm^{2} s^{-1}}$, $R=0.431\ \si{nm}$ (again, a spherical radius augmented by capture efficiency and spontaneous recombination distance considerations \cite{chiapetto2015nanostructure}) and a purely 3D-mobility. Here, the reference to "electron irradiation" come from the production of Frenkel pairs which are introduced progressively at $T=573\ K$ with a rate of $1\ \si{dpa s^{-1}}$. This rather extreme damage rate was chosen for numerical efficiency concerns rather than representativity in this first attempt. 

\begin{figure}
\includegraphics[width=0.7\textwidth]{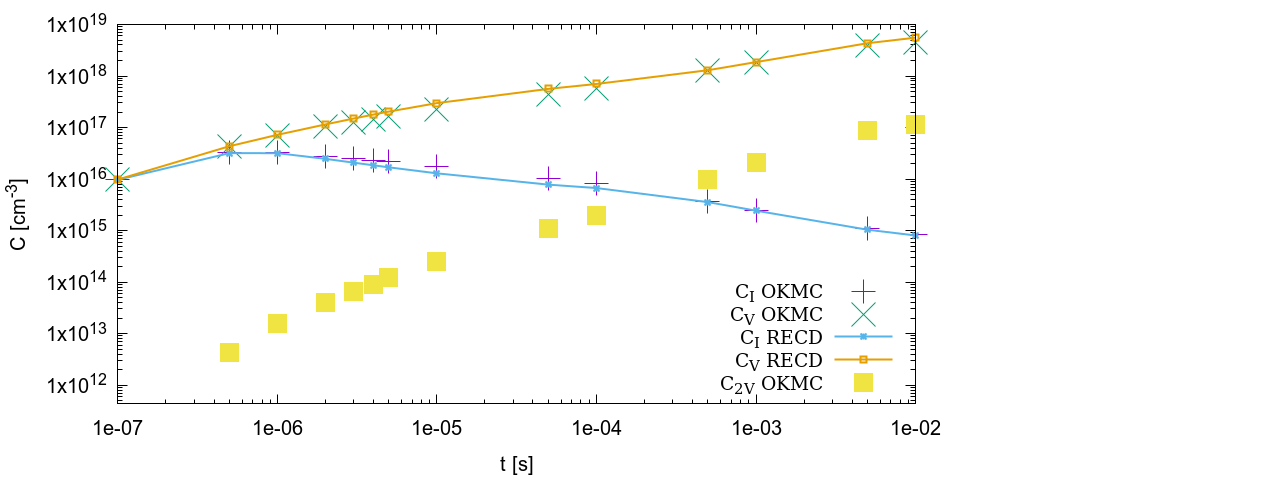}
\caption{Time evolution of the single interstitial, mono-vacancy and di-vacancy concentrations : $C_I$, $C_V$, $C_{2V}$ respectively (in $\si{cm^{-3}}$) under constant Frenkel pair production rate simulated by OKMC and RECD with the semi-analytical CSS.}
\label{C_IV_ARTICLE}
\end{figure}

\begin{figure}
\includegraphics[width=0.5\textwidth]{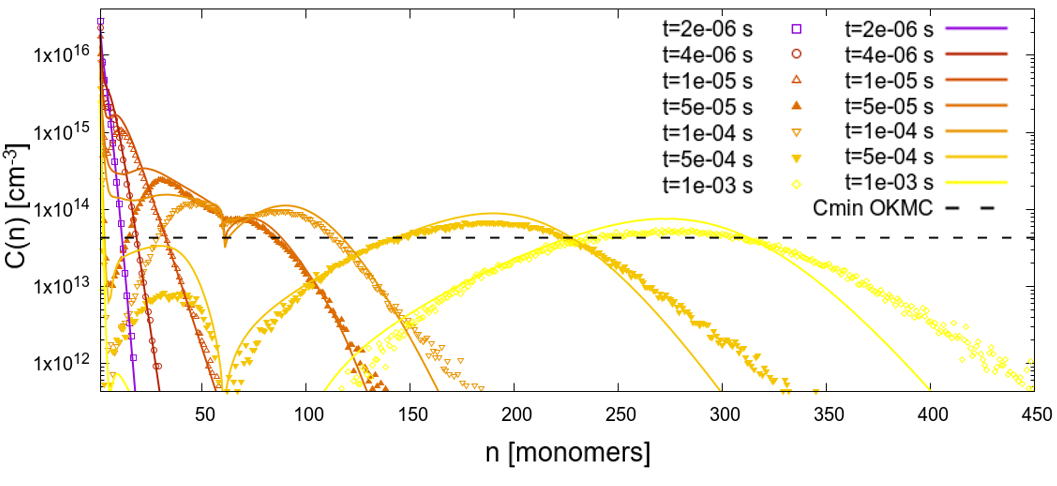}
\caption{Comparison of SIA clusters size distributions obtained by OKMC and RECD at different irradiation times $t$.}
\label{CirradIV}
\end{figure}

\begin{figure}
\includegraphics[width=0.5\textwidth]{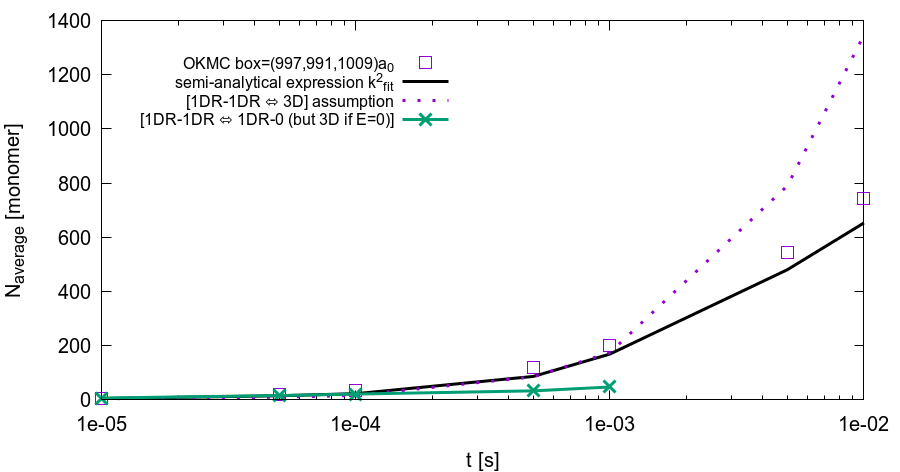}
\caption{Time evolution of the average size (in monomers) associated with the cluster distributions. OKMC is compared to RECD considering either the current CSS development and the two main CSS assumptions in the literature: the $\left[1DR-1DR \Leftrightarrow 3D-0\right]$ assumption and the $\left[1DR-1DR \Leftrightarrow 1DR-0\right]$ assumption.}
\label{Nave_irrad}
\end{figure}

The OKMC simulations consists in the average of one thousand runs of nine days longs simulations with box dimensions $(997,991,1009)a_{0}$. With the short irradiation times considered here, vacancies are too slow to observe any significant vacancy clustering: as can be seen in Fig. \ref{C_IV_ARTICLE}, not even di-vacancies ($C_{2V}$) exist in significant proportions compared to monovacancies ($C_{V}$). Free vacancies nevertheless play the most essential part of their role for kinetics of SIA clusters: SIA recombinations with vacancies limits SIA clusters growth. Detailed SIA cluster distributions are represented in Fig. \ref{CirradIV} for both OKMC and RECD. The time evolution of first moment of these distributions, $N_{\text{average}}$, is represented in Fig. \ref{Nave_irrad} and it is compared to the RECD with CSS assumptions that were previously discussed. In spite of some discrepancies at long times either for small or for very large sizes, the agreement between OKMC and RECD with the new CSS expression is very reasonable. Also, as previously discussed, it is quite likely that the agreement would be further increased using even larger boxes. 

We finally note that, when considering the two alternative CSS assumptions from the literature, the conclusion on their applicability is quite similar to the previous application: the kinetics resulting from the $\left[1DR-1DR \Leftrightarrow 3D-0\right]$ equivalence assumption are way too fast and those from the $\left[1DR-1DR \Leftrightarrow 1DR-0\right]$ assumption are way too slow.

\section{Discussion: the relevance of mixed mobility sinks strength to dislocation loop evolution modeling}\label{Discussion}

Aiming at providing general CSS expressions, the rotation energies of clusters' Burgers vector were taken here as generic "black box" parameters. From a physical point of view, the Burgers vector rotation phenomenon in a crowdion-bundle seems to originate from the rearrangements individual crowdions, but the confirmation of this mechanism is limited in MD to the small clusters \cite{Osetsky2000,Gao}, and little is known about its specific attempt frequencies and related entropic contributions.
Even more problematically, the elastic interactions between the cluster and other sources from the medium should also play a role. Elastic models \cite{Wolfer,Okita} show that, when exposed to the elastic field of a dislocation, the loops' habit planes tend to rotate (which is different from the thermally activated rotation of the Burgers vector) to minimise elastic energy thus loosing their pure edge orientation \cite{Bako}. This primarily plays a role in the decoration of dislocation lines by elastically trapped loops atmospheres \cite{Wen}, but it is also important at the loop's scale since it results in a misalignment of the loops' normal with the Burgers vector. A consequence of this is that the Burgers vector's rotation probability may not always bear a simple parameterisation with fixed rotation energy that only depends on its size: it may also depend on the surrounding elastic fields, which could lower the Burgers vector's rotation energies \cite{Wen}. By construction, with a mean-field method like cluster dynamics, we can only hope to treat these types of spatial effects in a very indirect way and this would be at the cost of quite heavy modifications of the original formalism.

Another important phenomenon is the trapping of loops by chemical heterogeneities. Carbon and nitrogen, even at a few tens of ppm concentrations can have a huge impact on the mobility of loops in ferritic materials at moderate temperature. Cottrell atmospheres of carbon atoms may form in the tensile zone of dislocation loops thus lowering the effective mobility of loops, and in the presence of vacancies, vacancy-carbon complexes may form \cite{Anento} and trap dislocation loops for long times. The effective migration energies of loops are thus, significantly increased by a function of the trap-loop binding energy. In other words, free species (i.e. not trapped) are expected to be in lower number densities than trapped ones in many conditions, so, at first sight, one may believe that $1DR-1DR$ CSSs may play a secondary role in the microstructure kinetics compared to $1DR-0$ ones. But we shall keep in mind that:
\begin{itemize}
\item At high temperatures, trapping is less effective. There could even exist a cross-over between trapping and its opposite effect, the elastic confinement of loops by oversized solute atoms \cite{Hudson2005}.
\item The vicinity of fixed sinks (dislocation network and grain boundaries) are often depleted of traps. It is indeed in these regions that greater populations of small dislocation loops (in the form of TEM "black dots") can be seen in complex alloys with many solutes and impurities such as industrial RPV steels at moderate doses. This is a local but crucial effect as it conditions the local mobility of dislocation which is a key point for mechanical properties evolution under irradiation.
\item Even if loops spend most of the time trapped, clusters from the resulting smaller population of detrapped loops may react quite intensively and depending on the detailed conditions create a specific population.
\item At least from some point, the traps should be saturated and the new free defect clusters produced in cascades should be more likely to interact with other mobile clusters than to be trapped. In the parameters set taken as an example in this study (see table \ref{tableMalerba}) the single interstitials are considered as 3D-mobile, and, from size 3, they start having a mixed mobility. Note also, that in other systems, the mono-interstitials may be crowdions and exhibit pure 1D-motion \cite{Amino2016}. In all these cases the $1DR-1DR$ CSS must be taken into account.
\item As we have seen in section \ref{Closure}, even with six orders of magnitude difference for the diffusion coefficients of two interacting 1D-mobile species, the $1D-0$ CSS approximation is a serious underestimation of the effective CSSs, contrary to the adapted $1D-1D$ expression. We can be more quantitative for this last but not least point: a popular way of accounting for trapping, inspired by the developments of Krishan \al \cite{Krishan} and others, is to simply lower the effective diffusion coefficient by a factor $\exp\left(-E_{\text{trap}}/k_B T \right)$, where $E_{\text{trap}}$ is the binding energy between the cluster and the trap. Actually, this corresponds to a lower bound for the effective diffusion coefficient as the trapping efficiency is assumed to be ideal. At $573\ {\si K}$, a trapping energy around $0.68\ \si{eV}$ corresponds to the factor $10^{-6}$ reduction of effective mobility. So, below this threshold trapping energy, a weakly trapped cluster should definitely be considered as mobile with respect to free clusters of the same size, as the error resulting from treating the interaction with the $1D-0$ CSS would be large according to figure \ref{k2_calc_S_k2_2Dand1D_vs_D1_-S-_D2_FEW_CURVES_ZOOM}. There are several defect sizes complying with this condition in the parameterisation exemplified in the reference parameterisation \cite{chiapetto2015nanostructure}: these are the smallest clusters but also, the most mobile and thus the most influential on the kinetics.
\end{itemize}

With these considerations in mind, mobile clusters coagulation (by diffusion of both reaction partners) should, at least be considered as a channel for loop growth competing with Ostwald ripening, and could even overtake it in specific conditions. Note that Monte-Carlo \cite{chiapetto2015nanostructure} and RECD simulations \cite{Kohnert} from the literature evoked the central role of mutual mobility in reproducing experimental loop growth kinetics, but the latter simulations (using RECD) did not benefit from the present developments on CSS expressions and rather relied on direct CSS additivity assumptions (such as those tested in section \ref{comparison}) which are not validated.
In a nutshell, for a population of trapped clusters, Ostwald ripening will dominate only if the cluster's dissolution rate (related to binding energies) is higher than its the detrapping rate.
For the reference parameterisation, trapping energies of loops range from $0.17$ to $0.6\ \si{eV}$ below a cluster size threshold value \cite{Note6} and then $1.1\ \si{eV}$ above the threshold. The corresponding binding energies values are always much higher than trapping energies at each cluster size (binding energies typically range from about $0.9\ \si{eV}$ up to the formation energy of interstitial, both bounds being so large that emissions are negligible for almost all clusters sizes at moderate temperatures). So if the raw diffusion coefficients of interstitial clusters are comparable in terms of orders of magnitude to that of the monomer, then the loops will be more likely to be detrapped rather than dissolve and feed the Ostwald ripening process. Of course, we may not prematurely conclude that coagulation always prevails, because high diffusion also promotes higher effective rates of emission (absorption rates appear in emission rates), but we stress out that the competition between the two processes must be considered.
This sheds new light on the range of validity of some of the major growth and coarsening theories: in the Lifshitz-Slyozov-Wagner (LSW) formalism, both reaction channels are indeed considered but the formulation of agglomeration rates restricts its application to 3D mobilities. Since no expression for absorption rates involving two $1DR$ species with arbitrary diffusion coefficients was available, the applicability of general results on growth asymptotics to these cases was unclear. Now we see that, once proper corrections are applied, absorption rates between such species lead to \textit{apparent} second-order kinetics rates just as 3D ones, and classical results may be qualitatively valid.

\section{Summary and conclusions}

Because the CSS dependence on concentrations, radii, diffusion coefficient and rotation energies is unknown and non-linear, the complete fitting of CSSs on all of their variables may represent quite a formidable task: any decent grid would require millions of conditions, all of which have quite different convergence behavior as the estimated CSSs will span over several orders of magnitude. To tackle the problem of the large set of conditions, the proposed approach is the following: fixing the defect concentrations and the radii, we generated a first data set making a grid on the diffusion coefficient ratio and the rotation energies. Then, we combined the analytical expressions developed in the companion paper as limiting cases, thus allowing us to extend the expressions to arbitrary concentrations and radii, as was further checked by analysis of the residuals in validation sets. The transition between the limiting cases can be modeled by two combinations of sigmoids: the first combination of sigmoids is used to describe the transition between the CSS when $D_A=D_B$, and the second couple of sigmoids are intended to reproduce the presence of the quite distinct mobility domains (consistently with the limiting cases) which appear clearly when mapping the effective exponents of the diffusion ratio correcting factor. This term appears to be the most crucial one to capture the many orders of magnitudes over which the CSS varies when the diffusion coefficient ratio goes from one to small values ($\Delta$ going from $0$ to large values). We also looked for the point from which the slowest specie should actually be considered as immobile in terms of reaction rates. The CSS with respect to a fixed sink would become more relevant after this point. This limit happens to be surprisingly low (high in terms of $\Delta$), which stresses out the necessity to use the $1DR-1DR$ CSS fit up to very small diffusion ratios. This should also be the case for the most important defect cluster couples (those mostly produced by cascades and that actually drive the dynamics) in typical irradiation conditions and would still hold when effective diffusion coefficients are considerably lowered by trapping. The relevance of $1DR-1DR$ reactions is thought to be quite general: depending on detailed conditions and especially on temperature, coagulation of mobile clusters may compete with Ostwald ripening or supersede it.
Finally, in the application of the RECD implementation, an in-depth validation of the formula was allowed thanks to a massive set of OKMC simulations with large box sizes. In passing, this also emphasized the necessity to use very large boxes with this type of complex mobility OKMC parameterisations to mitigate the box volume effects. As in moderate stiffness cases, RECD numerical schemes allow for very large time steps, this opens the way to very long-term (decades of physical time) simulations of the microstructure evolution fully accounting for the mixed mobility defect clusters which are intrinsically out of reach of any event-based method such as KMC.

\section{Acknowledgements}
Dr. Lorenzo Malerba is warmly acknowledged for initially pointing out the necessity of considering the mixed mobilities in OKMC simulations, which initiated this work on their RECD counterpart, as well as Dr. Christophe Domain who additionally performed a benchmark with its own implementation of effective CSS calculations within LAKIMOCA. Dr. Jean-Paul Crocombette is thanked for his suggestion to think about a "phase-diagram"-like representation of the CSS evolution with the mobility parameters. Dr. Thomas Jourdan is also acknowledged for his collaboration on the implementation within the code CRESCENDO.

\section{Funding}
This project initially received funding from the Euratom research and training program 2014-2018 under grant agreement No 661913 (SOTERIA).


\appendix
\section{Sink strengths in the 3D anisotropic case}\label{section_Sink_strengths_in_the_3D_anisotropic_case}
In a realistic irradiated microstructure, some small vacancy clusters and interstitial clusters will be mobile and their populations interact intensively. As these two types of clusters generally have $3D$ and $1D$ mobilities respectively, the general expression of the $1D-3D$ CSS is also needed for a comprehensive RECD parameterisation. Also, considering some state-of-the-art parameterisations (section \ref{OKMCparam}), the dumbbell is assumed to have a $3D$ mobility and larger clusters a $1D$ one, so the $1D-3D$ CSS are also needed for these reaction pairs.
Just as in the case of $1D-1D$ in the companion paper \cite{Adjanor1}, we may directly derive this expression from equivalent diffusion problems. 
Here also, we start by examining the additivity assumption of diffusion tensors. This assumption holds more naturally than in the $1D-1D$ case because, in the present case, one of the reactants has a $3D$ mobility, so the geometric condition for interaction does not require any specific treatment (whatever the geometric configuration the relevant capture distance is always $R=R_{A}+R_{B}$). 
In this specific case, the rate for $A-B$ absorptions, with respective diffusion tensors $\mathbf{D_{A}}$ and $\mathbf{D_{B}}$, are then equivalent to the rates for a specie with diffusion tensor $\mathbf{D_{A}}+\mathbf{D_{B}}$ with respect to a fixed sink. This assumption made, the case of $A$ having an isotropic 3D-random walk while $B$ has $1D$ motion, is easily handled with the "3D-anisotropy" results that follow.

If we consider that the total diffusion tensor has three diagonal components $D_{x}=D_{y}=D_{\rho}$ and $D_{z}$, the absorption rate can be cast in analytical forms in the following three limiting cases: 
\begin{equation}
\frac{\partial C_{A}}{\partial t} = - C_{A} C_{B} {\overline D}  \notag
\end{equation}
\begin{numcases}{\times}
4 \pi \left(\frac{D_{\rho}}{D_{z}}\right)^{1/3} R/{\ln\left[2\left(D_{\rho}/D_{z}\right)^{1/2}\right]},\text{for $D_{z} \ll D_{\rho}$},\\
4 \pi R,\text{for $D_{z} \simeq D_{\rho}$},\\ 
8 \left( \frac{D_{z}} {D_{\rho}} \right)^{1/6} R, \text{for $D_{z} \gg D_{\rho}$}
\end{numcases}

where only the long-time terms were retained from the developments of Woo \cite{Woo2} and ${\overline D}=(D_{\rho}^2 D_{z})^{1/3}$ is the rescaled average diffusion coefficient relevant to this type of anisotropy. It is generally convenient in RECD to have absorption rates proportional to the sum of diffusion coefficients $D_{A}+D_{B}$ that we can state as $D_{z}$ and assume to be constant without loss of generality. Indeed, the diffusion coefficient dependencies of the previous cases can be cast in the following form:
\begin{equation}
{\overline D}\left(\frac{D_{z}}{D_{\rho}}\right)^\delta = {D_{z}}\left(\frac{D_{z}}{D_{\rho}}\right)^{\delta-\frac{2}{3}} 
\end{equation}

Now, to establish the CSS expression relevant to our case, we can choose $D_{\rho}=D_{B}$, which yields ${\overline D}\left(\frac{D_{z}}{D_{\rho}}\right)^\delta =D_{z}\left(\frac{D_{A}}{D_{B}}\right)^{-\frac{1}{2}}$ diffusion ratio dependency when $D_{A} \gg D_{B}$, thus exhibiting a characteristic exponent $(-1/2)$ for the diffusion coefficient ratio.


\section{Search for general simulation setup and convergence criteria} \label{Search}
\subsection{Technicalities of the simulation box size setup}
Because 1D-mobility may be far less efficient than the 3D one to sample the space (as reflected considering roughly $k^2_{1D-0} \propto k^2_{3D-0} \Phi$ \cite{Adjanor1}, $\Phi$ being a volume fraction) very large simulation boxes may be required. As Malerba and co-workers pointed out \cite{Malerba}, assigning the boxes' dimensions to three different prime numbers is sufficient to prevent a mobile with $\langle111\rangle$ glide direction to stay confined in the diagonal or any periodic path (as it happens when having periodic boundary conditions in a box whose dimensions have common divisors). Nevertheless, at variance with the authors' case where only sink concentration dependence was studied varying the box size containing one fixed sink (of type $B$) in a box of volume $V$, in the present case we perform more general calculations where many mobile clusters are present in the calculation. In that case, it may not be necessary to choose "non-cubic" boxes (as opposed to "quasi-cubic" ones with $L_x \simeq L_y \simeq L_z$ like, for example, a $(991 \times 997 \times 1009)$ box would be, in lattice parameter $a_{0}$ units). 

\subsection{Criteria for the convergence of CSS estimates}
\subsubsection{First criterion: compatibility between the box size and the number of clusters}\label{firstCriterion}
The previous condition is necessary but not sufficient to ensure correct estimations of the CSS. The box size should be carefully chosen so that there is at least one interacting pair along the shortest box dimension (linear densities of pairs times the shortest box dimension is greater than one). Adopting a slightly more probabilistic approach: assuming $L_x \le L_y \le L_z$, the probability of having more than two interacting species along one of the ${L_y L_z}$ segments of dimensionless length $L_x$ \cite{Note2} is
\begin{eqnarray}
P(N_t \ge 2)&=&1-P(N_t=0)-P(N_t=1)  \\
&=&1-(1-p)^{L_x}-L_x p (1-p)^{L_{x}-1}  \\
&\simeq& p^2 L_x^2,  \label{binomial}
\end{eqnarray}
assuming a binomial distribution of A species along $L_x$.
Here, $p$ is the probability for one site to be occupied by a $A$-type particle:
\begin{eqnarray}
p=C_A {a_{0}^3}=\frac{N_{A}}{L_x L_y L_z}, 
\end{eqnarray}
is assumed to be small \cite{Note3}.

Noting that there is, on average, 
\begin{eqnarray}
\frac{N_{A}}{L_y L_z}
\end{eqnarray}
$A$-type particles along one $L_x$ segment, one should then sum up the probability Eq.\ref{binomial} over all the ${L_y L_z}$ segments of length $L_x$ (the probabilities for the individual segments are assumed to be independent) for a criterion on having at least one $L_x$ segment with two A-particles aligned. Note that, as stated before, $A-A$ reactions are not accounted for, and the criteria we look for in fact concerns $A-B$ interactions. But it turns out to be more simple to manipulate only one concentration $C_A$, and then extract a conservative criterion assuming $C_A \le C_B$.
Doing so, we end up with the following crudely approximated but conservative criterion 
\begin{eqnarray}
(p L_x )^2 L_y L_z \simeq 1. \label{crit}
\end{eqnarray}
If, more realistically, one wishes to account for the minimum distance between species to be greater than twice the capture radius (clusters are now placed on a grid now with spacing $R$ instead of $a_{0}$ and $p$ is now $C_A R^3$),
\begin{eqnarray}
N_A^{\text{min}}\gtrsim L_x {\frac{a_{0}} {R}},
\end{eqnarray}
with $C_A \le C_B$ and $L_x \le L_y \le L_z$, this gives the order of magnitude for the minimum number of species to place in the box to have a large proportion of $A-B$ reactions. Note that this criterion should be conservative also regarding glide direction effects because it should be even more easily met when glide directions are non-collinear to the boxes' directions as, in the case of $\langle111\rangle$-glide, for example, interaction probabilities might be enhanced by the shift of linear trajectories after crossing a periodic boundary. 

\begin{figure}
\includegraphics[width=0.5\textwidth]{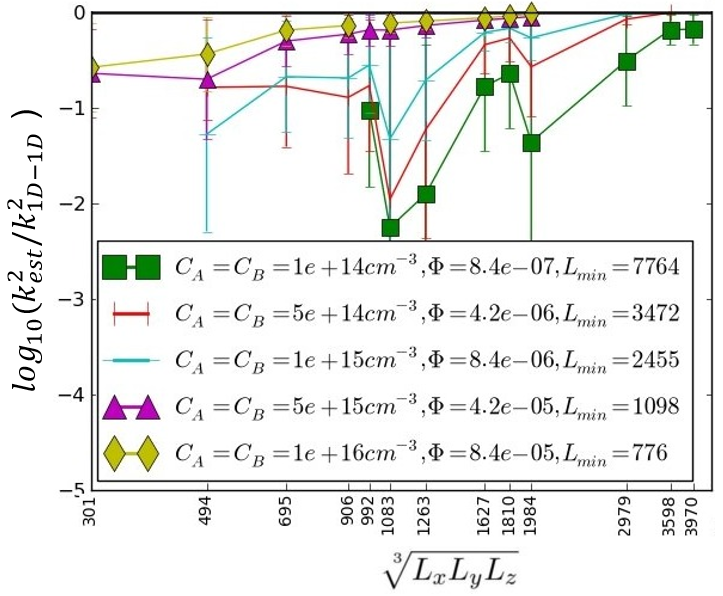}
\caption{The decimal logarithm of the ratio of the estimated CSS over the analytical value ($\log_{10} \left(\sfrac{k^2_{\text {est}}}{k^2_{1D-1D}}\right)$) as a function of $\sqrt[3]{L_x L_y L_z}$ for various $(C_{A},R)$ couples ($R_{A}=R_{B}=1\ \si{nm}$). Points of a given curve all correspond to the same computation times. Significant convergence is reached only when the box size reaches approximately the $L_{\text {min}}$ criterion. }
\label{boxSize}
\end{figure}

This criterion happens to be well-validated if we consider the results of Fig.~\ref{boxSize}. On this graph are plotted the decimal logarithm of the ratio of CSS estimates from OKMC simulation over the analytical expression $k^2_{1D-1D}$ \ref{Eq_k2_1D-1D}. In these simulations (with the parameters $E_{A} = E_{B} =2\ \si{eV}$, $D_{A}=D_{B}$, $R_{A}=R_{B}=1\ \si{nm}$), the analytical expression and simulation estimates of the CSS should match at least within the uncertainty of the estimates (i.e. the $\log_{10}$ of their ratio should reach $0$ within the error bars in Fig.~\ref{boxSize}). This ratio is given as a function of the average box size $\sqrt[3]{L_x L_y L_z}$, so a logarithm of the ratio far from zero denotes the inability of the simulation setup to reproduce correct sink strength values due to small box size effects which are notably sharp in the case of 1D-mobility simulations. On the caption of this figure are also displayed the values of the minimal box sizes for convergence:
\begin{equation}
L_{\text {min}}=\frac{1}{\sqrt{C_{A} R^3}}
\end{equation} 
for each concentrations $C_{A}=C_{B}$. The $L_{\text {min}}$ values are obtained applying preceding criterion Eq. \ref{crit} but for a fixed concentration (for this criterion $L_x \simeq L_y \simeq L_z$ is assumed). We see that this simplistic criterion catches reasonably well the order of magnitude of the minimal box size for the onset of convergence: below these values, the convergence is not reached regarding standard deviations of the estimates. 

\subsubsection{Second criterion: relation between the number of clusters and the minimal number of reactions}
The first criterion ensured us that, for a given concentration, the box will be large enough so that all A-clusters have at least one B-type partner for a potential reaction. Another criterion is needed on the extent of the sampling so that most of A and B-clusters have at least one heterotypic reaction.
To simplify this condition this quite demanding condition, the sampling of reactions should be extended enough to have at least as many absorptions recorded as the total number of mobiles. This is because absorption rates should reflect all contributions to the reaction probabilities: on one hand, rare reactions occurring by very fast 1D collision of clusters which randomly happen to be initially placed very close to each other, and, on the other hand, contributions from the vast majority of slow (due to the inefficient 1D-sampling of space) non-collinear 1D-1D reactions. Stopping the run before having significant chances that all mobiles had at least one interaction, and restarting it with another initial random configuration of defect would result in sampling only the fast reactions and thus considerably overestimating the CSS. A direct consequence is that it limits the use of "naive parallelisation" \cite{Note4} for the longest calculation: each run must be continued until most clusters of both types have reacted. This point is also very important and if it is overlooked it can be a major source of inaccuracy in CSS estimations: the CSS may vary by several orders of magnitudes going from fast components to its slowest ones.

\subsubsection{Third criterion: averaging estimates over several runs}
Applying the second criterion with a given initial configuration, one produces a single estimate of the CSS by averaging the reaction time for a number of reactions greater than the total number of mobiles. Now, from the statistical point of view, one should estimate the coefficient of variation (the ratio of the standard deviation to the mean) related to estimates.
As a consequence, one should sample several tens of initial configurations and judge whether the sampling is sufficient to have a coefficient of variation.

\section{Formulation of the semi-analytical fitting function for the CSS isosurfaces} \label{appendix_fitting_cube}
\begin{figure}
\includegraphics[width=0.5\textwidth]{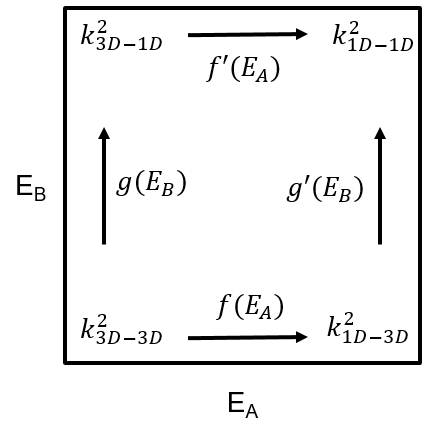}
\caption{Schematic view of the relation between the fitting function $f$, $f'$, $g$ and $g'$ and the limiting cases for the CSS when $D_{A}=D_{B}$ in the $(E_{A},E_{B})$ plane.}
\label{delta-scheme0}
\end{figure}

Let us now examine the fitting strategy adopted to reproduce the main features of the isosurface. It consists in using several functions to describe the transitions between known CSS expressions as limiting cases. 
As sketched in Fig.~\ref{delta-scheme0}, when $D_{A}=D_{B}$, one very general formulation of the CSS may be:
\begin{align} \label{MostGeneralKappa}
k^2(E_{A}, E_{B}) =& f(E_{A})\left( \left( 1-g(E_{B}) \right) k^2_{1D-3D} + g(E_{B}) k^2_{1D-1D} \right) \\ \notag
&+ \left(1- f'(E_{A})\right) \times\left( \left( 1-g'(E_{B})\right) k^2_{3D-3D} + g'(E_{B}) k^2_{3D-1D} \right)
\end{align}
with the functions $f$, $f'$, $g$ and $g'$ having the constraints: $f(0)=f'(0)=g(0)=g'(0)=0$ and $f(1)=f'(1)=g(1)=g'(1)=1$ so that the four limiting cases are recovered:
\begin{eqnarray} 
&&k^2(0, 0) = k^2_{3D-3D}\left(D_{A}, D_{A}\right) \\ 
&&k^2(0, \infty) = k^2_{3D-1D}\left(D_{A}, D_{A}\right) \\ 
&&k^2(\infty, 0) = k^2_{1D-3D}\left(D_{A}, D_{A}\right) \\ 
&&k^2(\infty, \infty) = k^2_{1D-1D}\left(D_{A}, D_{A}\right) 
\end{eqnarray}

If we impose that $E_{A}$ and $E_{B}$ play symmetric roles ($k^2(E_{A}, E_{B}) =k^2(E_{B}, E_{A})$, when $D_{A}=D_{B}$) and use $k^2_{3D-1D}\left(D_{A}, D_{A}\right)=k^2_{1D-3D}\left(D_{A}, D_{A}\right)$, then we must have $f=g$ and $f'=g'$, and we can cast the CSS to the following form:
\begin{eqnarray}
k^2(E_{A}, E_{B}) &= & f(E_{A})\left( \left( 1-f(E_{B}) \right) k^2_{1D-3D} + f(E_{B}) k^2_{1D-1D} \right) \\ \notag
&+&\left( 1-f'(E_{A}) \right)\left( \left( 1-f'(E_{B})\right) k^2_{3D-3D} + f'(E_{B}) k^2_{1D-3D} \right)\label{generalKappa}
\end{eqnarray}
In practice, experimenting with various fitting possibilities, it turns out that we can make this form even simpler taking:
\begin{eqnarray}
f'(x)=f(x) = \sigma(\lambda_f,\varepsilon_f,x) 
\end{eqnarray}
and a simple fitting procedure leads to $\lambda_f=8$ and $\varepsilon_f=0.2$, with overall 5\% discrepancy with the simulated data set and a maximum deviation of about 15\% for the worse point.

Following the previous findings, we will now assume that the fit can be extended to $D_{A}>D_{B}$ scaling it with the ratio of diffusion coefficients to a power that depends on the dimensionality of the mobilities:
\begin{equation}
k^2\left(E_{A}, E_{B}, D_{A}, D_{B}\right) = k^2(E_{A}, E_{B}) \times \left(\frac{D_{A}}{D_{B}}\right)^{\delta(E_{A}, E_{B})} 
\end{equation}
We are now left with the task of defining the exponent function $\delta$. The values of $\delta$ that should be matched according to the analog limiting cases are sketched in Fig.~\ref{delta-scheme}:
\begin{itemize}
\item $(E_{A}=0,\ E_{B}=0)$ corresponds to $3D-3D$ absorption rates which simply depend on $(D_{A}+D_{B})$, that makes $(1+(D_{A}/D_{B})^{-1})$ when factoring for $D_B$ the CSS, which is close to $1=(D_{A}/D_{B})^0$ when $D_{A} \gg D_{B}$ in other words a close to zero $\delta$ value,
\item $(E_{A}=\infty,\ E_{B}=\infty)$ corresponds to $1D-1D$ absorption rates with a characteristic exponent of $-1/3$, consistently with the related limiting case \cite{Adjanor1},
\item $(E_{A}=\infty,\ E_{B}=0)$ similarly corresponds to $1D-3D$ with $D_{A}>D_{B}$ absorption rates leading to a $-1/2$ exponent \ref{section_Sink_strengths_in_the_3D_anisotropic_case}.

\begin{figure}
\includegraphics[width=0.5\textwidth]{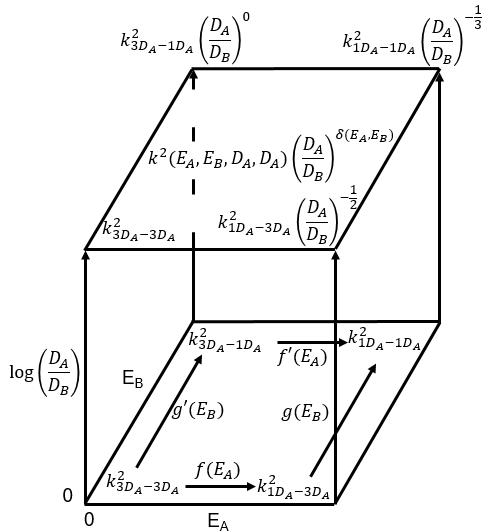}
\caption{Schematic view of the analytical expressions for the limiting cases (corresponding to the evolution of the CSS along the vertical edges of the cube) for CSSs in the $(E_{A},E_{B}, \Delta)$ space. The functions $f$, $f'$, $g$ and $g'$ were shown to be identical by symmetry and practical arguments.}
\label{delta-scheme}
\end{figure}

\end{itemize}

\begin{figure}
\includegraphics[width=0.5\textwidth]{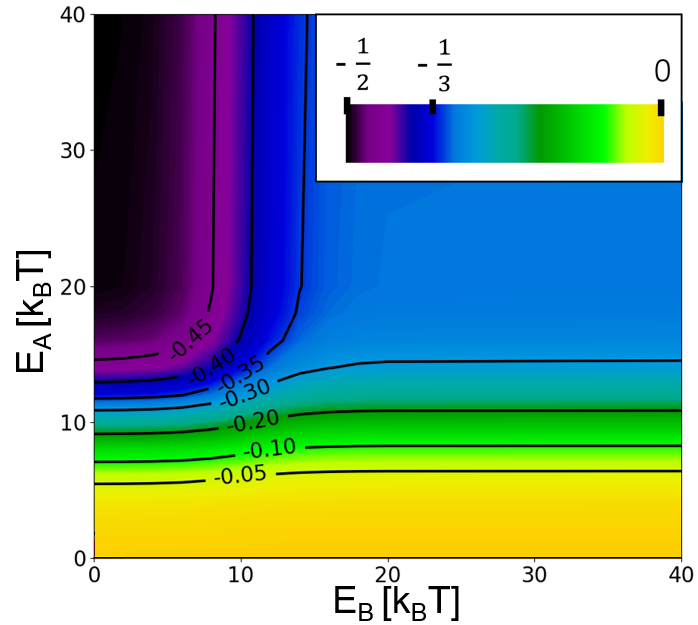}
\caption{Isolines representation of the exponent function $\delta$ Eq.\ref{delta}.}
\label{effective_delta_function_map}
\end{figure}

One very simple choice for an exponent function $\delta$ meeting these requirements is:
\begin{equation}
\delta(E_{A}, E_{B}) = \sigma(\varepsilon,\lambda, E_{A}) \left(-\frac{1}{2}+\frac{1}{6} \sigma(\varepsilon',\lambda', E_{B})\right). \label{delta}
\end{equation}
The first sigmoid multiplies the whole expression allowing to reproduce a smooth transition from the 3D to 1D mobility of A species, which should indeed be in the leading term since they have the largest diffusion coefficient ($D_{A}\ge D_{B}$). The second sigmoid just accounts for the behavior of the B-species which are less mobile, as reflected by the weaker influence of this term. The isolines of this function are represented in Fig.~\ref{effective_delta_function_map} and exhibit a $3D-3D$ domain (yellow), a $1D-1D$ domain (blue) and a $1D-3D$ (purple) domain consistently with effective exponent calculations in section \ref{semiAnalytical}.

Performing a basic fitting procedure for this function $\delta$ leads to $\lambda=10$ and $ \varepsilon=0.5$ for both sigmoid functions. 
Note that the $\delta_{\text{eff}}$ map analysis (Fig.~\ref{logarithmic_derivate} and Eq. Fig.~\ref{effective_delta_function_map}) shows us that trying to improve the fit would require a more elaborate expression of the function $\delta$ than Eq.\ref{delta}, because it should also depend on $\Delta$ (as discussed in section \ref{Closure}).

\section{Temperature and lattice dependent semi-analytical formula} \label{extension_to_other_temperatures}
\begin{figure}
\includegraphics[width=0.5\textwidth]{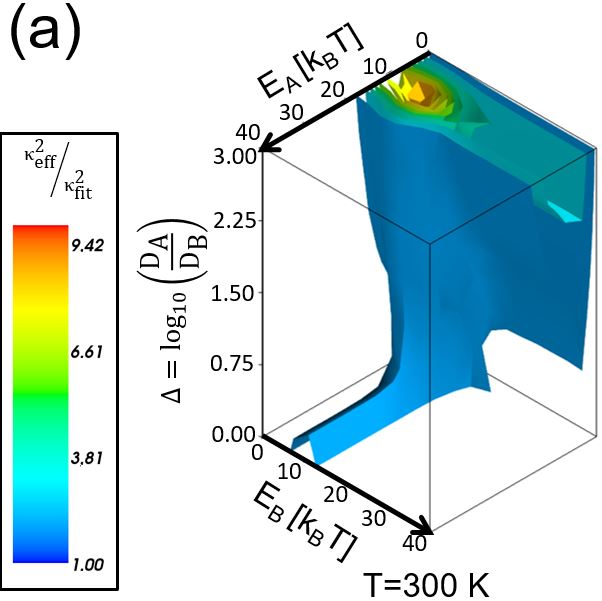}
\includegraphics[width=0.5\textwidth]{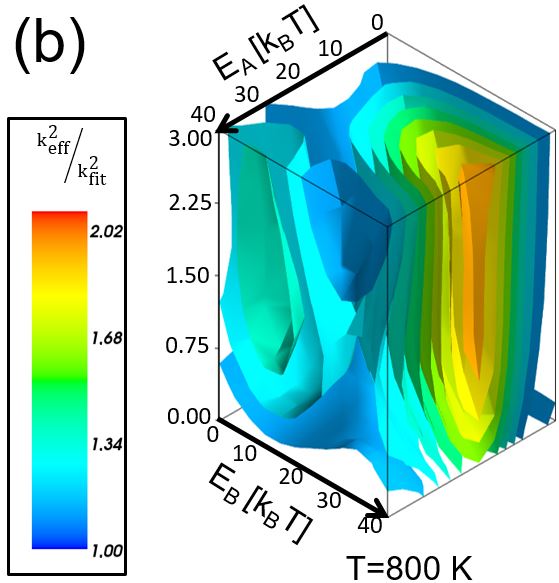}
\caption{Isosurfaces of the ratios $k^2_{\text{eff}}/k^2_{\text{\text{fit}}}$. (a) $T=300\ {\si K}$, (b) $T=800\ {\si K}$, and $C_{A}=C_{B}=\num{5e16} \si{cm^{-3}}, R_{A}=R_{B}=1\ \si{nm}$ in both cases. The vertical axis represents $\Delta$.}
\label{quadratic_residue_2000p3_C5e16_300K-800K}
\end{figure}

In the semi-analytical expression of the CSS, the transition functions between well-defined analytical CSS expressions were fitted on OKMC calculation for a given lattice type, a given first nearest neighbor distance, and at a given temperature. To generalize it, we may first note that the temperature only plays a role in the mean-free path before rotation $\ell$ as the formula is directly parameterised on diffusion coefficients. It may then be adapted to depend on the couple $(\ln(\ell_A), \ln(\ell_B))$, with
\begin{equation}
\ln(\ell)=\frac{E}{2 k_B T} + \ln d_j,
\end{equation} 
instead of rotation energies $(E_{A}, E_{B})$. Also as the $1DR-0$ CSS limiting cases are not explicitly accounted, only the sigmoid transition functions must be adapted:
\begin{equation}
\sigma'(\ell, \lambda,  \varepsilon, T) = \frac{1}{1+\exp \left[ - { 2 \lambda k_B T_0} \left(\ln(\ell) - \ln(d_j) -\frac{ \varepsilon}{2 k_B T}\right)\right]},
\end{equation}
which ensures $\sigma'(\ell, \lambda,  \varepsilon, T_0) = \sigma(E, \lambda,  \varepsilon)$.
In fact, this formulation is useful only when adapting to both a different temperature and a different lattice (reflected by jump distance $d_j$). When only adapting to a different temperature, the transformation may be reduced to substituting $E$, $\varepsilon$ and $\lambda$ to $E/T$, $\varepsilon/T$, $\lambda/T_{0}$ (with $T_{0}=573\ {\si K}$) respectively in the original expressions of the sigmoids. The discrepancies related to the absence of $1DR-0$ terms in the fit and the odd parity issue of the sigmoid function could possibly be worsen when applying to other temperatures, so the quality of the adapted fit at $T=300\ {\si K}$ and $T=800\ {\si K}$ was studied. The isosurfaces of the ratios $k^2_{\text{eff}}/k^2_{\text{\text{fit}}}$ are shown in Fig.~\ref{quadratic_residue_2000p3_C5e16_300K-800K}. The situation is almost the opposite in the two cases. For the low temperature, the discrepancy gets higher but in an even more restricted zone than at the reference temperature $T_{0}$, so the overall accuracy is not significantly changed (about 18 \%). For the higher temperature, the maximum discrepancy is lower than for the reference, but it is more widespread over the parameter space, and the location of the maximum discrepancy (about 2) is different. This also leads to a global accuracy of the same order as for the initial fitting temperature (about 23 \%). This also suggests using temperature-dependent terms in the transition functions for applications requiring much higher levels of accuracy.

\section{Degree one homogeneity of the semi-analytical formula} \label{homogeneity}
Here, we wish to assess the degree one homogeneity property of $k^2_{\text{fit}}$:
\begin{equation}
k^2_{\text{fit}}(C_A, C_B)=k^2_{\text{fit}}(C, \lambda C)\simeq \lambda \cdot k^2_{\text{fit}}(C,C).
\end{equation} 
Because the $k^2_{3D}$ components of $k^2_{\text{fit}}$ trivially respects it, it only remains to verify this property for $k^2_{1D-1D}$. For simplicity, let us only consider the concentration-dependent part of the absorptions rates ${\mathcal A}$ (diffusion coefficients are here omitted):
\begin{equation}
{\mathcal A} (C_A, C_B)= \frac{C_A C_B}{\ln\left(\frac{1}{2}(C_A+C_B) R^3\right)}={\mathcal A} (C_B, C_A).
\end{equation} 
Now, without any loss of generality, we can assume that $C$ designates the concentration of the majority specie, say $C=C_A$, and that the minority specie is then $C_B=\lambda C$, with $\lambda \le 1$.
\begin{equation}
{\mathcal A} (C_A, C_B)={\mathcal A} (C, \lambda C)= \frac{\lambda C^2}{\ln\left(\frac{1}{2}(C+\lambda C) R^3\right)}=\frac{\lambda C^2}{\ln\left(\frac{1+\lambda}{2}\right)+\ln\left(\frac{3}{4 \pi} \Phi\right)}, 
\end{equation} 
where $\Phi$ is $\frac{4 \pi}{3} R^3 C$ relates to the volume fraction of the majority species.
Now, since: 
\begin{equation}
{\mathcal A} (C, C)= \frac{C^2}{\ln\left(\frac{3}{4 \pi} \Phi\right)}, 
\end{equation}
assuming the degree 1 homogeneity property consists in doing the following approximation:
\begin{equation}
{\mathcal A} (C, \lambda C)= \frac{\lambda C^2}{\ln\left(\frac{1+\lambda}{2}\right)+\ln\left(\frac{3}{4 \pi} \Phi\right)} \simeq \frac{\lambda C^2}{\ln\left(\frac{3}{4 \pi} \Phi\right)}=\lambda {\mathcal A} (C, C), 
\end{equation}
Let us further justify it. In the denominator $\left[{\ln\left(\frac{1+\lambda}{2}\right)+\ln\left(\frac{3}{4 \pi} \Phi\right)}\right]$, the $\ln\left(\frac{3}{4 \pi} \Phi\right)$ term is dominant because of the volume fraction-related term inside of it. Assuming $\lambda=1/2$ to have concrete evaluations in the table \ref{tableHomogeneity}, we have $\ln\left(\frac{1+\lambda}{2}\right)=\ln\left(\frac{3}{4}\right)\simeq -0.29$. This is negligible compared to $\ln\left(\frac{3}{4\pi} \Phi\right)$, evaluated at second line of the table, even up to extreme values $\Phi=0.1$ (10\% volume fraction of the main species), hardly met in typical applications.
In terms of Taylor expansion, the approximation made reads more precisely as:
\begin{eqnarray}
{\mathcal A}(C, \lambda C) &=& \lambda C^2 X \left[1-\ln\left(\frac{1+\lambda}{2}\right)  X + o(X) \right] \\
&=& \lambda {\mathcal A}(C,C) \left[1-\ln\left(\frac{1+\lambda}{2}\right)  X + o(X) \right], 
\end{eqnarray}
where $X=\frac{1}{\ln\left(\frac{3}{4 \pi} \Phi\right)}$. The evaluation of $\left[1-\ln\left(\frac{1+\lambda}{2}\right)  X \right]$  for the $\lambda=1/2$ is given is the last line of the table. The conclusion is that the assertion ${\mathcal A}(C, \lambda C) = \lambda {\mathcal A}(C,C)$ is valid within 90\%-95\% precision or more because the first-order correction term $\left[\ln\left(\frac{1+\lambda}{2}\right) X\right]$ is always small in realistic applications.

\begin{table}
\begin{tabular}{|c|c|c|c|c|c|c|}
\hline
$\Phi$           & $0$     &  $10^{-6}$     &   $10^{-4}$    &   $10^{-1}$     & $0.1$       & $1$  \\
\hline
$\ln\left(\frac{3}{4 \pi} \Phi\right)$     & $-\infty$     &  $-15.2$     &   $-10.6$    &   $-6.0$     & $-3.7$       & $0$  \\
\hline
$X=\frac{1}{\ln\left(\frac{3}{4 \pi} \Phi\right)}$     & $0$     &  $-0.06$     &   $-0.09$    &   $-0.16$     & $-0.26$       & $-\infty$  \\
\hline
$\frac{\ln\left(\frac{3}{4 \pi} \Phi\right)}{\ln\left(\frac{1+\lambda}{2}\right)+\ln\left(\frac{3}{4 \pi} \Phi\right)}$    &  $ $ & $0.981$     &  $0.973$     &   $0.954$    &   $0.928$     &    $ $      \\
\hline
$\left(1-\ln\frac{3}{4}X\right)$     &    $ $  &  $0.981$     &   $0.972$    &   $0.952$     & $0.922$       & $ $  \\
\hline                                                                     
\end{tabular}                                                              
\caption{Numerical evaluation of the approximations for degree one homogeneity. \label{tableHomogeneity}}  
\end{table}

\end{document}